\documentclass[11pt]{article}

\usepackage[authoryear,longnamesfirst]{natbib}

\usepackage[margin=1in]{geometry}
\usepackage[T1]{fontenc}
\usepackage{lmodern}
\usepackage{xspace}
\usepackage{pifont}
\usepackage{xurl}
\usepackage[hidelinks]{hyperref}
\usepackage{setspace}

\usepackage{graphicx,verbatim}
\usepackage{booktabs}       
\usepackage{amsfonts}       
\usepackage{amsmath,amsfonts,amssymb}
\usepackage{cleveref}
\usepackage{pdfpages}
\usepackage[textfont=it]{caption}
\usepackage{adjustbox}
\usepackage{float}
\usepackage{bm} 
\usepackage[inline]{enumitem} 
\usepackage{lineno}

\newcommand{\Ab}{\ensuremath{\text{A}\beta}\xspace}
\newcommand{\bTh}{\ensuremath{\boldsymbol{\theta}}}
\newcommand{\bPhi}{\ensuremath{\boldsymbol{\phi}}}
\DeclareMathOperator{\diag}{diag}

\title{LNODE: latent dynamics reveal the shared spatiotemporal structure of amyloid-$\beta$ progression}                      

\author{%
Zheyu Wen\textsuperscript{1,*}
\qquad
George Biros\textsuperscript{1,2,*}
\\[0.6em]
for the Alzheimer's Disease Neuroimaging Initiative
and the A4 Study Team
\\[0.9em]
\parbox{0.88\textwidth}{%
\centering\small
\textsuperscript{1}Oden Institute for Computational Engineering and Sciences,\\
The University of Texas at Austin, Austin, Texas 78712, USA
\\[0.35em]
\textsuperscript{2}Walker Department of Mechanical Engineering,\\
The University of Texas at Austin, Austin, Texas 78712, USA
\\[0.7em]
\textsuperscript{*}Corresponding authors:
\href{mailto:zheyw@utexas.edu}{zheyw@utexas.edu};
\href{mailto:biros@oden.utexas.edu}{biros@oden.utexas.edu}
}}

\date{}

\begin{document}
\maketitle
\begin{abstract}
We present and analyze LNODE, a mechanism-based phenomenological model for amyloid beta (\Ab) dynamics, calibrated using positron emission tomography (PET) imaging. \Ab is a key biomarker of Alzheimer's disease. 
LNODE is designed to support the fusion, harmonization, quantitative analysis, and interpretation of \Ab PET scans. 
We evaluate LNODE on 1461 subjects in the ADNI cohort and 1,070 subjects in the A4 Study, using MUSE and DKT anatomical atlases. 
LNODE is formulated as a regional neural ordinary differential equation (ODE) model that is jointly calibrated on \emph{all} available scans within a cohort. 
The model captures the spatial propagation, proliferation, and clearance of \Ab and incorporates a latent-state representation that modulates \Ab dynamics. 
The temporal evolution of these latent states is governed by cohort-shared parameters, enabling LNODE to represent both population-level trajectories and subject-specific deviations. 
The proposed model demonstrates strong parameter identifiability and stability properties, supported by synthetic experiments and analysis of the Hessian condition number.
To mitigate overfitting and reduce spurious correlations, LNODE is intentionally underparameterized, employing approximately five to ten parameters per subject. 
Despite this parsimonious parameterization, LNODE achieves cohort $R^2 > 0.99$ in both the ADNI and A4 datasets. 
LNODE exhibits strong predictive performance: in the A4 cohort, it accurately forecasts the \Ab PET signal in previously unseen follow-up scans, including cases with inter-scan intervals exceeding four years.
Clustering in the learned latent-state space reveals distinct subgroups, consistent with the existence of different subtypes of Alzheimer's disease progression.
\end{abstract}



\noindent\textbf{Keywords:}
\Ab computational modeling;
Latent network ODE;
Biomarker prediction;
\Ab subtype


\section{Introduction}
Alzheimer's disease is a highly prevalent neurodegenerative disorder and the leading cause of dementia, characterized by progressive impairments in memory, cognition, and behavior.
Key imaging biomarkers in biological staging include amyloid  and tau PET. 
Indeed, convergent evidence indicates that aberrant \Ab deposition begins decades before the onset of obvious clinical symptoms~\citep{villemagne2013amyloid}. For example, the five-year risk of progression to mild cognitive
impairment (MCI) or dementia was 57\% in \Ab positive/tau-positive individuals, compared to  6\% in \Ab-negative / tau-negative individuals~\citep{moscoso2025frequency}. 
Despite criticisms of the role of amyloid in AD, \Ab remains one of the hallmark biomarkers and characterizes the progression of the disease~\citep{kepp2023amyloid,frisoni2022probabilistic}.

PET with the AV45 radiotracer enables in vivo imaging of the spatial distribution and longitudinal evolution of \Ab pathology in the human brain~\citep{wong2010vivo}. 
Rich datasets such as the Alzheimer's Disease Neuroimaging Initiative (ADNI) dataset~\citep{petersen2010alzheimer} and Anti-Amyloid Treatment in Asymptomatic Alzheimer's Disease (A4)~\citep{sperling2014a4} enable the design of computational models for  the quantitative interpretation of \Ab-PET scans to  improve diagnostic precision, prognostic assessment, and subject stratification.

Some groups use statistical inference and machine learning methods to analyze these AD imaging datasets \citep{gong2026taugennet,wang2025learning,danuncover}. 
A model takes diffusion-reaction framework and replaces an explicit nonlinear reaction terms with an operator learning approach. The method uses graph Laplacian eigenfunctions as a basis and learns their coefficients to model personalized AD biomarkers~\citep{wang2025learning}.
Another work models tau spreading as a mean field game with a variational form equivalent to a diffusion-reaction model, leading to a forward-backward saddle-point problem solved via Generative Adversarial Networks and demonstrated on a cortical surface mesh with over 100k vertices~\citep{danuncover}.

A complementary approach to statistical interpretations is to use a  mechanism-based phenomenological model using Markov processes, ODEs, and other forms of discrete or continuous dynamical systems and calibrate it using imaging data~ \citep{young2024data}.
Indeed, several groups have developed  such  models and used them for discovering and predicting spatio-temporal patterns in \Ab deposition dynamics~\citep{jack2013tracking,vogel2020spread,vogel2023connectome,chaggar2025personalised,dore2013cross,young2018uncovering}.
Examples of results from such studies include staging of \Ab and its relationship to cortical thickness~\citep{dore2013cross,mattsson2019staging}, the estimation of changes in biomarker burden over time~\citep{chaggar2025personalised}, and the characterization of disease subtypes~\citep{vogel2021four}.

Mechanism based models  usually  have a small number of parameters and thus, easier to train.
However, we also have several challenges, which we summarize here in the context of our approach. 
\begin{enumerate*}
\item {\bf Time scales:}
  Amyloid deposition dynamics evolve over decades, while the time interval between imaging sessions can be less than a year for some subjects.
  This makes calibration of the model difficult, since the changes between subsequent visits are small and often dominated by noise.
  Indeed, in \citep{wen2025single}, we showed that reconstructing dynamics with such observations is not robust. 
\item {\bf Initial deposits (or seeds):}
  An alternative formulation is to assume much longer time scales that  include the initiation of the disease.
  This approach allows for a more stable reconstruction~\citep{wen2025single,wen2025aligning} but involves additional unknowns: the location of the initiation of \Ab.  In our approach, we opt for reconstructing the initial seeding, thus recovering the \emph{entire unobserved history} of the disease dynamics. 
  Such trajectories can be used to shed light on the disease dynamics that eventually may lead to new therapies.

\item {\bf Inter-individual variability of time scales:}
  The different rates of progression and spatio-temporal patterns across individuals suggest that the  use of chronological age to track time can be problematic.
  For these reasons, several papers~\citep{young2018uncovering,vogel2021four} have used the notion of a \emph{disease age} that we also adopt here~\citep{ghazi2021robust}.
\item {\bf Unknown dynamics:}
  The misfolding of \Ab starts in the extracellular space and eventually accumulates and aggregates to form plaques.
  However, there are no known upscaled dynamics and the precise mechanisms are not fully understood~\citep{chen2017amyloid,mohamed2016amyloid}. 
  For this reason, we and others have opted for  phenomenological models that try to capture the basic mechanisms of accumulation and spatial spread.
\item {\bf PET signal interpretation:}
  There is no universal definition on how to connect the PET signal to amyloid deposition.
  In addition to the standardized uptake ratio (SUVR),  there are mixture models of abnormal \Ab,  global measures of \Ab, such as the centiloid scale, and others
  \citep{klunk2015centiloid,vogel2020spread}.
  We prefer to use the distributional measure of the regional abnormality of \Ab because we have found that it outperforms the other measures~\citep{wen2025single}
\item {\bf Brain parcellation:} Spatial analysis is typical of done using regions of interest (ROIs) to account for the noisy PET signal.
 ROIs are related to functional parcellations. We test LNODE on two different parcellations commonly used in AD studies.
\item {\bf Multiple biomarkers:} Phenomenological models are able to integrate multiple biomarkers  that, in addition to amyloid,  account for atrophy, glucose uptake, and tau tangles~\citep{vogel2021four}.
In particular, neurofibrillary tau protein tangles (or simply tau) represent another  major biomarker of Alzheimer's disease~\citep{franzmeier2020functional} 
Here,  we focus exclusively on \Ab scans. In the past, we have applied similar multimodal scan methods, for example, for tau, \Ab, and atrophy~\citep{wen2025aligning}. In the present study, we focus specifically on \Ab  because the ADNI dataset comprises substantially more longitudinal \Ab PET scans than tau PET scans.  Future  extensions of this work will incorporate tau pathology and atrophy.  
\end{enumerate*}

\paragraph{Contributions}
In~\cite{wen2025lnode}, we introduced the main features of LNODE and examined its ability to fit existing data. (Additional discussion on related work can be found in~\cref{s:discussion}.) Our main contribution here is to carefully analyze the well-posedness of inverting for the  parameters of this model, evaluate its performance on a much larger dataset from ADNI and extend it to the A4 study, and test the ability of the method to predict \emph{unseen scans}. 

LNODE comprises a set of ODEs for abnormal and normal \Ab, and a number of  latent states.  Mechanisms included in LNODE include spatial spreading, proliferation, glymphatic clearance, and coupling to latent states. To avoid over-parameterization, our model has  cohort-shared parameters and subject-specific parameters.
The cohort-shared parameters govern the latent-state dynamics through a graph neural network ordinary differential equation formulation. 
We introduce a formulation and an algorithm that reconstructs these parameters by solving a cohort-level inverse problem across all subjects.
Sparse selection over latent states enables flexible modeling of subject-specific \Ab dynamics, allowing individualized trajectories to emerge from shared underlying progression patterns.

In summary, the goal of this study is to answer the following questions:
\begin{enumerate}
\item Is the inverse problem of fitting LNODE parameters to data numerically stable? How does it depend on the cohort size and the number of hidden states?
  (see Fig.~\ref{fig:invertibility_hessian})
\item Compared with widely used computational models of \Ab propagation, can the proposed LNODE framework achieve substantially improved reconstruction and prediction accuracy at both individual and cohort levels?
\item How does LNODE's performance varies across the ADNI and A4 datasets? How do they depend on the ROI definition?
\item Do the estimated parameters shared by the cohort generalize to accurately predict unseen \Ab scans for subjects undergoing potential diagnostic transitions?
\item Do the learned latent states uncover clinically significant subtypes of \Ab progression between subjects?
\end{enumerate}  

\section{Methods}\label{s:method}

\begin{figure}[!htbp]
\centering
\includegraphics[width=1\textwidth]{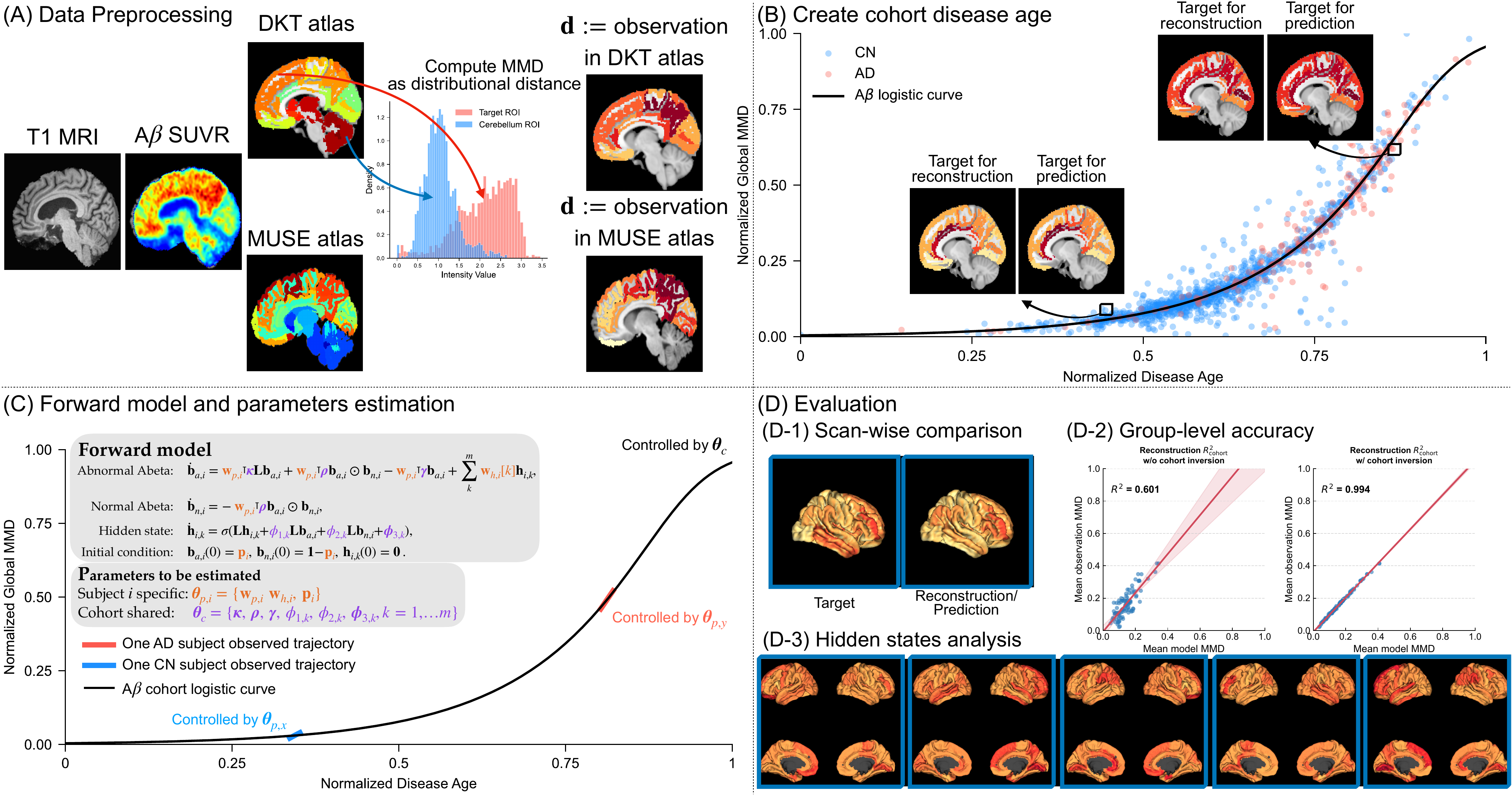}
\caption{
  \textbf{Overview of the proposed LNODE method.} 
  The workflow comprises four main components:
  \textbf{(A) Data preprocessing.} We collect T1-weighted MRI and \Ab PET data from the ADNI and A4 Study datasets. 
  PET images are co-registered to each subject's first T1-weighted MRI scan, and the brain is parcellated into \( N \) regions of interest using either the MUSE or DKT atlas. 
  For each scan, we extract a regional \Ab burden, on scalar value per ROI.
  This value is computed using the maximum mean discrepancy metric, which measures the distributional distance between voxel-wise SUVR values in a cortical ROI and those in the whole cerebellar gray matter reference region. 
  The resulting regional MMD values serve as the observational ground truth for the \Ab scans.
  \textbf{(B) Disease age creation.} We map the chronological age of each scan to the corresponding global \Ab MMD value using an affine transformation followed by a logistic function. 
  The resulting normalized disease age \( t \in [0,1] \) is used as the time variable for modeling \Ab progression.
  \textbf{(C) LNODE model.} We model \Ab progression using a latent neural ordinary differential equation (LNODE) framework. 
  The latent state dynamics are governed by a single-layer graph neural network, while the \Ab burden dynamics are modeled using a diffusion--reaction--clearance system that incorporates sparse selection over the latent states. 
  The model includes cohort-shared parameters to capture common \Ab progression patterns across subjects, as well as subject-specific parameters to account for individual variability.
  \textbf{(D) Model training and evaluation.} The model is trained by minimizing the scan-wise mismatch across the cohort \Ab data. 
  Model performance is evaluated in terms of scan-wise reconstruction and prediction accuracy, cohort-level accuracy, and the interpretability of the learned latent states.
}
\label{fig:method}
\end{figure}

\subsection{Image acquisition and processing.}
Fig.~\ref{fig:method} provides an overview of the proposed LNODE framework.
Fig.~\ref{fig:method}(A) illustrates the data preprocessing steps. 
We obtain clinical T1-weighted MRI and amyloid-beta (\Ab) PET data from two datasets: the Alzheimer's Disease Neuroimaging Initiative (ADNI) and the Anti-Amyloid Treatment in Asymptomatic Alzheimer's (A4) Study. 
We include subjects who have both T1-weighted MRI and \Ab PET scans available. 
In total, we include 1461 ADNI subjects, and 1070 A4 Study subjects.
The data preprocessing pipeline follows the standard procedures described by ADNI. 
Specifically, PET scans are first motion-corrected using the AFNI program 3dvolreg, after which the motion-corrected frames are averaged to generate a single volumetric PET image per scan. 
The longitudinal PET images are co-registered to each subject's baseline T1-weighted MRI using FSL~\citep{smith2004advances}. 
We perform Skull stripping on the T1-weighted MRI scans using FSL. 
The resulting brain masks are applied to the co-registered PET images.
Subsequently, PET images are intensity normalized using the whole cerebellar gray matter as the reference region to compute SUVR maps. 
Finally, the brain is parcellated into \( N \) ROIs using two atlases: the MUSE atlas and the DKT atlas~\citep{doshi2016muse,klein2012101}. 
Our analysis focuses on cortical \Ab dynamics, where the MUSE atlas defines 98 cortical ROIs and the DKT atlas defines 62 cortical ROIs.
The proposed algorithm is validated using both atlases. 
Results presented in the main text are based on the DKT atlas, while complementary results using the MUSE atlas are reported in the supplementary material.
 
\subsection{Regional abeta-PET data preprocessing.}
For each scan, we extract a single scalar value from each ROI to represent the regional \Ab burden. 
Specifically, for each cortical ROI, we collect the voxel-wise SUVR values within the ROI and quantify the distributional discrepancy between these values and those from the reference whole cerebellar gray matter region.
We use the  maximum mean discrepancy, which is a kernel-based statistical distance measure, to quantify the degree of difference between two probability distributions~\citep{gretton2012kernel}. 
A Gaussian kernel is used in the MMD computation, and kernel variance is selected to maximize MMD.
This computation is performed independently for each subject's ROI on both the MUSE and DKT atlases.
The resulting regional MMD values serve as the observational ground truth for the \Ab scans.
 
\subsection{Computation of disease age for each scan.}
Fig.~\ref{fig:method}(B) illustrates the computation of the disease age.
It follows the procedure described in~\citep{ghazi2021robust,wen2025aligning}.
Briefly, we define an affine transformation with unknown subject-specific slope and bias parameters that maps the chronological age of each scan to an unnormalized disease age.
We assume that this can be mapped to the global \Ab MMD load via a logistic function with cohort-shared unknown parameters.
Using this assumption, the subject-specific and cohort-shared parameters are jointly estimated by minimizing the data mismatch between the observed global \Ab MMD values (summation of regional MMD values) and the logistic model prediction values across all subjects and scans. 
The estimated disease ages are normalized using min-max normalization.
This procedure aligns all scan data onto a unified disease progression timeline. 
Each scan is consequently assigned a normalized disease age \( t \in [0,1] \), which is used as time variable in the subsequent LNODE modeling.
 
\subsection{Modeling \Ab dynamics with latent states.}
Fig.~\ref{fig:method}(C) illustrates the proposed LNODE model.
The proposed LNODE model is designed to predict subject-specific \Ab dynamics while simultaneously capturing common disease progression patterns shared across subjects. 
Accordingly, the model incorporates both cohort-shared parameters and subject-specific parameters.
Cohort parameters govern population-level disease dynamics and subject-specific parameters explain individual variability.
Let \( N \) denote the number of gray matter ROIs. 
We model the dynamics of three types of variables: the abnormal amyloid-beta burden \( \mathbf{b}_a(t) \in \mathbb{R}^N \), the normal amyloid-beta burden \( \mathbf{b}_n(t) \in \mathbb{R}^N \), and a set of latent state variables \( \mathbf{h}_k(t) \in \mathbb{R}^N \), for \( k = 1, \ldots, m \). 
Here, \( t \) denotes disease age.
Motivated by biological evidence that amyloid-beta propagates between cells through neuronal connections, we model the spreading dynamics of abnormal \Ab using a diffusion--reaction--clearance system of ordinary differential equations (ODEs). 
In this formulation, diffusion captures the propagation of \Ab along neuronal connections, reaction models the conversion of normal \Ab into abnormal \Ab, and clearance represents the removal of abnormal \Ab.
Similar model structural forms have been widely used in the literature to model tau or \Ab propagation~\citep{thompson2020protein,vogel2020spread,raj2025understanding,wen2025single}.
To capture additional complex dynamics that cannot be adequately described by the classical diffusion--reaction--clearance framework, we introduce latent state variables \( \mathbf{h}_k(t) \). 
The dynamics of these latent states follow a graph neural network (GNN) formulation, which jointly models the diffusion of \( \mathbf{h}_k(t) \), \( \mathbf{b}_a(t) \), and \( \mathbf{b}_n(t) \) over the brain network. 
The abnormal \Ab state \( \mathbf{b}_a(t) \) performs sparse selection over the latent states, thereby defining a subject-specific baseline \Ab dynamics. 
The normal \Ab state \( \mathbf{b}_n(t) \) represents the pool of amyloid that can be converted into abnormal \Ab through interactions with \( \mathbf{b}_a(t) \).
\par
Following standard modeling approaches for \Ab propagation, we represent inter-regional interactions using a graph Laplacian. 
Specifically, using tractography data from a reference atlas, we construct a sparse ROI connectivity matrix \( \mathbf{W} \), and define the (negative) graph Laplacian as
\[
\mathbf{L} := \mathbf{W} - \operatorname{diag}\!\left(\sum_{s \neq r} [\mathbf{W}]_{rs}\right).
\]
With these definitions, our LNODE model for subject $i$ reads as follows:
\begin{subequations}
\label{eq:forward}
\begin{align}
  \text{Abnormal Abeta:}\quad & \dot{\mathbf{b}}_{a, i} = \mathbf{w}_{p, i}^\intercal\boldsymbol\kappa \mathbf{L}_i \mathbf{b}_{a, i} + \mathbf{w}_{p, i}^\intercal\boldsymbol\rho \mathbf{b}_{a, i}\odot\mathbf{b}_{n, i} - \mathbf{w}_{p, i}^\intercal\boldsymbol\gamma \mathbf{b}_{a, i}\notag\\ 
                              & \quad\quad\quad\quad      + \sum_{k}^m\mathbf{w}_{h, i}[k]\mathbf{h}_{i,k},\\
  \text{Normal Abeta:}\quad & \dot{\mathbf{b}}_{n, i} = - \mathbf{w}_{p, i}^\intercal\boldsymbol\rho \mathbf{b}_{a, i}\odot\mathbf{b}_{n, i},\\
  \text{Latent state:}\quad & \dot{\mathbf{h}}_{i,k} = \sigma(\mathbf{L}_i\mathbf{h}_{i,k} + \phi_{1, k}\mathbf{L}_i\mathbf{b}_{a, i} + \phi_{2, k}\mathbf{L}_i\mathbf{b}_{n, i} + \boldsymbol\phi_{3,k}),\\
  \text{Initial condition:}\quad & \mathbf{b}_{a, i}(0)=\mathbf{p}_{i},\ \mathbf{b}_{n, i}(0)=\mathbf{1} - \mathbf{p}_{i},\ \mathbf{h}_{i, k}(0)=\mathbf{0}.
\end{align}
\end{subequations}
where $\dot{\mathbf{b}}$ indicates time derivative. $\mathbf{w}_{p, i}\in\mathbb{R}^o$ are designed to do sparse selection of candidate parameters $\boldsymbol\kappa$, $\boldsymbol\rho$ and $\boldsymbol\gamma\in\mathbb{R}^o$. 
$\mathbf{w}_{h, i}\in\mathbb{R}^k$ ``selects'' sparse dominant latent states, and $\odot$ represents the element-wise product.
In the latent-state dynamics ROI coupling is done through $\mathbf{L}_i$ for $\mathbf{h}_{i,k}$, $\mathbf{b}_{a, i}$,  $\mathbf{b}_{n, i}$; 
$\sigma: \mathbb{R}^{N}\to\mathbb{R}^{N}$ is an element-wise nonlinear function (e.g. ReLU). 
In equation~(\ref{eq:forward}), the abnormal \Ab $\mathbf{b}_{a,i}$ is coupled to the latent states through the $\mathbf{w}_{h, i}[k]\mathbf{h}_{i, k}$.
\par
In the literature, it is a common practice to use a fixed, population-averaged and precomputed graph Laplacian for the term $\mathbf{L}_i=\mathbf{\bar{L}}$, where $\mathbf{\bar{L}}$ is a sparse and constant matrix.
Here, we introduce a learnable graph Laplacian that is both individualized and disease progression dependent, enabling the modeling of possible dynamic connectivity changes over time. 
The temporal graph Laplacian is defined as:
\begin{subequations}
\label{eq:laplacian}
\begin{align}
    \text{Attention-based Connectivity:} \quad & \mathbf{A}_i(t) = \text{softmax}[(\mathbf{w}_{Q}\mathbf{b}_{a, i}(t)^\intercal)(\mathbf{w}_{K}\mathbf{b}_{a, i}(t)^\intercal)],\\
    \text{Symmetrization:} \quad & \mathbf{W}_{i}(t) = (\mathbf{A}_i(t) + \mathbf{A}_i^\intercal(t)) / 2, \\
    \text{Graph Laplacian:} \quad & \mathbf{L}_{i}(t) = \mathbf{\bar{L}} + (\mathbf{W}_{i}(t) - \text{Diag}(\mathbf{W}_{i}(t)\mathbf{1})),
\end{align}
\end{subequations}
where $\mathbf{w}_{Q},\ \mathbf{w}_{K}\in\mathbb{R}^N$ are \emph{cohort-shared learnable parameters}, i.e., every subject in the cohort has the same $\mathbf{w}_Q$ and $\mathbf{w}_K$. 
Further, we make $\mathbf{L}_i$ subject-specific and time-varying by including the $\mathbf{b}_{a,i}(t)$ terms. 
This formulation encodes, for each ROI, the relative influence of all other ROIs on \Ab propagation at time $t$ for subject $i$. 
To maintain parameter efficiency, we use a single attention layer. 
The softmax function is applied to normalize the relative influence scores between ROIs into a probability distribution. 
Subsequently, we symmetrize the resulting matrix $\mathbf{A}_i(t)$ and derive its graph Laplacian for integration into our latent ODE model.
As the attention weights are shared across the entire cohort, the number of free parameters per subject is small. 
In the paper, we test both constant graph Laplacian and attention inspired graph Laplacian.
\par
Finally, the inversion parameters are split to subject-specific parameter set $\bTh_p$ and cohort-shared parameter set $\bTh_c$:
\begin{align*}
\bTh_p &:= \left\{\mathbf{w}_{p, i}, \mathbf{w}_{h, i}, \mathbf{p}_{i}\ i=1,\ldots,n, k=1,\ldots,m\right\},\\
\bTh_c &:=  \left\{\boldsymbol\kappa, \boldsymbol\rho, \boldsymbol\gamma, \phi_{1, k}, \phi_{2, k}, \bPhi_{3, k}, \mathbf{w}_{Q}, \mathbf{w}_{K}\ k=1\ldots m\right\},\\
\bTh   &:= \left\{ \bTh_p, \bTh_c\right\}. 
\end{align*}

\subsection{Cohort inversion.}
We use Gauss-Newton method to solve the following inverse problem to reconstruct the parameters $\bTh$.
We define $\mathbf{d}_{ij}$ as MMD extracted from the $j_{\mathrm{th}}$ scan of subject $i$, acquired at disease age $t_{ij}$. 
Also, we use $s_i$ to denote the number of scans for subject $i$. 
We use the following objective function to estimate parameters $\bTh$:
\begin{equation}\label{e:objective}
\!\!\!\mathcal{J}=\sum_{i}^{n} \frac{1}{2s_i}\sum_j^{s_{i}}\left\|\mathbf{b}_{a, i}(t_{ij}) - \mathbf{d}_{ij}\right\|_2^2 + \lambda_1 \sum_{i}^n\left\|\mathbf{w}_{h, i}\right\|_1 + \lambda_2 \sum_{i}^n\left\|\mathbf{w}_{p, i}\right\|_1,
\end{equation}
where $\left\|\cdot\right\|_1$ represents the $\ell_1$ norm to promote sparsity in parameter and latent state selections;
\par
With these definitions, the overall inverse problem reads as follows:
\begin{equation}\label{eq:inv}
  \begin{split}
    & \min_{{\bTh_p>0, \bTh_c}} \mathcal{J}(\bTh) \text{ subject to:}\\
    &\qquad \text{ equation~\ref{eq:forward} holds and }
    \left\|\mathbf{p}_{i}\right\|_0=s^{\text{max}}\ \forall i.
 \end{split}
\end{equation}
Note that we set $\ell_0$ norm on IC to enforce a sparse initial condition for \Ab. 
For the sparsity constraint of hidden states selection, we propose to also test a more rigorous math formulation as the follows:
\begin{equation}\label{e:inv2}
  \begin{split}
    & \min_{{\bTh_p>0, \bTh_c}} \mathcal{J}(\bTh)=\sum_{i}^{n} \frac{1}{2s_i}\sum_j^{s_{i}}\left\|\mathbf{b}_{a, i}(t_{ij}) - \mathbf{d}_{ij}\right\|_2^2 + \lambda_2 \sum_{i}^n\left\|\mathbf{w}_{p, i}\right\|_1, \text{ subject to:}\\
    &\qquad \text{ equation~\ref{eq:forward} holds and }
    \left\|\mathbf{p}_{i}\right\|_0=s^{\text{max}}\ \text{ and } \sum_{i}^n\left\|\mathbf{w}_{h, i}\right\|_0=1 \forall i.
 \end{split}
\end{equation}
We use gradient method to solve equation~(\ref{eq:inv}).
To compute the gradient with respect $\bTh_p$ and $\bTh_c$ we use a Lagrangian/adjoint formulation, in which we first solve the backward-in-time adjoint ODEs and then we accumulate the gradients. 
More details about how to solve the optimization problem can be available in~\citep{wen2025lnode}.
We initialize the subject-specific parameters $\bTh_{p}$ by zeros and the cohort-shared parameters $\bTh_{c}$ by random values drawn from $[0, 1]$.
 
\subsection{Connectivity matrix.}
The present study validates the proposed LNODE model using two different atlases: MUSE and DKT. 
Each atlas uses a distinct structural connectivity matrix \( \mathbf{W} \) to represent the anatomical connections between ROIs.
The construction of the structural connectivity matrix for the MUSE atlas follows the procedure described in~\citep{wen2025single}. 
Briefly, a simulated concentration distribution is initialized in a source ROI, and an anisotropic diffusion partial differential equation (PDE) is used to model the propagation of the concentration through the white matter. 
The diffusion direction and intensity are governed by Diffusion Tensor Imaging (DTI) data from healthy subjects.
The amount of concentration reaching a target ROI is then used to define the connectivity strength between the source and target ROIs. 
This procedure is repeated for all pairs of ROIs to construct the full connectivity matrix. 
The MUSE connectivity matrix is derived using diffusion tensor imaging (DTI) data from 20 healthy subjects in the Harvard Aging Brain Study (HABS)~\citep{dagley2017harvard}.
For the DKT atlas, the anatomical connectivity matrix is adopted directly from the construction described by Vogel \emph{et al.}~\citep{vogel2020spread}.
 
\subsection{Hyper-parameters selection.}
The number of latent states \( m \) and the dimension of the parameter vectors \( \boldsymbol{\kappa} \), \( \boldsymbol{\rho} \), and \( \boldsymbol{\gamma} \) (denoted by \( o \)) are two hyperparameters of the proposed forward LNODE model. 
The number of latent states \( m \) controls the model capacity to capture complex \Ab dynamics. 
We evaluate the model using different values of \( m \), specifically \( m = 0, 1, 3, 5, 10 \). 
Larger values of \( m \) enable the model to represent more complex dynamics but also increase the risk of overfitting. 
When \( m = 0 \), the model reduces to a classical diffusion--reaction--clearance model without latent states.
We also investigate different values of \( o \), with \( o \in [1, 10] \), and observe no performance improvement when \( o > 1 \). 
Consequently, we set \( o = 1 \) in all experiments to reduce model complexity.
\par
The sparsity regularization parameters \( \lambda_1 \) and \( \lambda_2 \) in equation~(\ref{e:objective}) are hyperparameters of the inversion algorithm. 
The parameter \( \lambda_1 \) controls the sparsity of the latent state selection weights \( \mathbf{w}_{h,i} \), while \( \lambda_2 \) controls the sparsity of the candidate parameter selection weights \( \mathbf{w}_{p,i} \).
We evaluate different values of \( \lambda_1, \lambda_2 \in \{0.001, 0.01, 0.1, 0.2\} \). 
Based on the prediction tasks performance and the sparsity of the estimated parameters, we set \( \lambda_1 = 0.1 \) and \( \lambda_2 = 0.01 \).
\par
We impose bounded constraints on all parameters during optimization. 
Specifically, the subject-specific parameters in $\bTh_{p}$ are restricted to lie within $[0,1]$, as $\mathbf{w}_{h,i}$ and $\mathbf{w}_{p,i}$ represent selection weights and $\mathbf{p}_i$ denotes the initial \Ab burden. 
For the cohort-shared parameters, $\boldsymbol{\kappa}$, $\boldsymbol{\rho}$, and $\boldsymbol{\gamma}$ are constrained to $[0,20]$ based on prior estimates from diffusion--reaction models. 
In addition, we bound $\phi_{1,k}$, $\phi_{2,k}$, and $\boldsymbol{\phi}_{3,k}$ within $[-10,10]$ to prevent parameter saturation at the boundaries.
 
\subsection{Evaluation metrics.}
Fig.~\ref{fig:method}(D) summarizes the key evaluation objectives of this study, including individual-level and cohort-level model accuracy as well as the analysis of learned latent states.
We choose a few metrics to quantify the reconstruction accuracy.
We use the $R^2\in[-\infty, 1]$ score to quantify the proportion of variance in the observed data that the model explains.
We compute two $R^2$ variants: a per scan $R_{\text{scan}}^2$; and a cohort-level score $R_{\text{cohort}}^2$. 
Specifically, for subject $i$ and scan $j$, we define $R_{\text{scan}, ij}^2$ as
\[
R_{\text{scan}, ij}^2(\mathbf{d}_{ij}, \mathbf{b}_{a, i}(t_{ij})) = 1 - \frac{\sum_{\ell=1}^N\left(\mathbf{d}_{ij}[\ell] - \mathbf{b}_{a, i}[\ell](t_{ij})\right)^2}{\sum_{\ell=1}^N\left(\mathbf{d}_{ij}[\ell] - \bar{d}_{ij}\right)^2},
\]
where $\ell$ represents the $\ell^{\text{th}}$ ROI and $\bar{d}_{ij}=\frac{1}{N}\sum_{\ell=1}^N\mathbf{d}_{ij}[\ell]$.
For $R_{\text{cohort}}^2$, we denote $\mathbf{d}_{\text{avg}}=\frac{1}{n}\sum_{i=1}^n\frac{1}{s_i}\sum_{j=1}^{s_i}\mathbf{d}_{ij}$ as averaged observation across scans and subjects, 
and $\mathbf{b}_{a, \text{avg}}=\frac{1}{n}\sum_{i=1}^n\frac{1}{s_i}\sum_{j=1}^{s_i}\mathbf{b}_{a, i}(t_{ij})$.
Then,
\[
R_{\text{cohort}}^2(\mathbf{d}_{\text{avg}}, \mathbf{b}_{a, \text{avg}}) = 1 - \frac{\sum_{\ell=1}^N\left(\mathbf{d}_{\text{avg}}[\ell] - \mathbf{b}_{a, \text{avg}}[\ell]\right)^2}{\sum_{\ell=1}^N\left(\mathbf{d}_{\text{avg}}[\ell] - \bar{d}_{\text{avg}}\right)^2},
\]
We also define the $\ell_2$ norm relative error to quantify the parameter per scan reconstruction error.
Specifically, for parameter $\theta$, we define the relative error as
\[e_{\bm\theta} = \frac{\left\|\hat{\bm\theta} - \bm\theta_{\text{true}}\right\|_2}{\left\|\bm\theta_{\text{true}}\right\|_2},\]
where $\hat{\bm\theta}$ is the reconstructed parameter and $\bm\theta_{\text{true}}$ is the ground truth.
For observation scan $\mathbf{d}_{ij}$, we define the relative error as
\[e_{\text{scan}, ij} = \frac{\left\|\mathbf{b}_{a, i}(t_{ij}) - \mathbf{d}_{ij}\right\|_2}{\left\|\mathbf{d}_{ij}\right\|_2},\]
where $\mathbf{b}_{a, i}(t_{ij})$ is the reconstructed scan and $\mathbf{d}_{ij}$ is the ground truth.
In addition, we evaluate the stability of the estimated parameters under different random initializations by 
\[\Xi_{\bm\theta} = \frac{1}{R} \sum_{r=1}^R \frac{\left\|\bm\theta_r - \bar{\bm\theta}\right\|_2}{\left\|\bar{\bm\theta}\right\|_2},\]
where $\bar{\bm\theta} = \frac{1}{R}\sum_{r=1}^R \bm\theta_r$ is the mean of the estimated parameter from $R$ different random initializations.

\subsection{Invertibility analysis.}
\label{subsec:invertibility}
We analyze the invertibility of the model by computing the condition number of the Hessian matrix of the objective function. 
We define the linearized simplified version of the abnormal \Ab dynamics as:
\begin{subequations}
  \label{eq:linearized forward}
\begin{align}
\frac{\partial \mathbf{b}_i}{\partial t} &= (\mathbf{w}_i^\intercal\bm\kappa) \mathbf{L} \mathbf{b} + (\mathbf{w}_i^\intercal\bm\rho) \mathbf{b} + \sum_j w_{ij} \mathbf{h}_{i,j}, \\
\frac{\partial \mathbf{h}_{i, k}}{\partial t} &= \mathbf{L} \mathbf{h}_{i, k} + \phi_1 \mathbf{L}\mathbf{b}_i + \bm\phi_3, \\
\mathbf{b}_i(0) &= \mathbf{p}_i, \quad \mathbf{h}_{i, k}(0) = \mathbf{0}, \quad k = 1, \dots, s.
\end{align}
\end{subequations}
where $\mathbf{L} \in \mathbb{R}^{N \times N}$ is symmetric positive semi-definite with eigendecomposition $\mathbf{L} = \mathbf{Q}\bm\Lambda \mathbf{Q}^\intercal$, $\bm\Lambda = \diag(\lambda_1, \dots, \lambda_N)$, $\lambda_i \ge 0$. 
The unknown parameters are $\bm\theta = (\mathbf{p}_i, \mathbf{w}_i, \bm\kappa, \bm\rho, \phi_1, \bm\phi_3)$, where $\mathbf{p}_i \in \mathbb{R}^N$ is the initial condition, $\bm\kappa > 0$ is the diffusion coefficient vector, and $\bm\rho > 0$ is the growth rate vector.
$\phi_1$ and $\bm\phi_3$ are cohort-shared parameters.
The data $\mathbf{d}_i \in \mathbb{R}^N$ is the measurement of $\mathbf{b}_i$ at time $T > 0$.
We do the following transformation: $\hat{\mathbf{b}} = \mathbf{Q}^\intercal \mathbf{b}$, $\hat{\mathbf{d}} = \mathbf{Q}^\intercal \mathbf{d}$, $\hat{\mathbf{p}} = \mathbf{Q}^\intercal \mathbf{p}$, $\hat{\mathbf{h}}_{i,j} = \mathbf{Q}^\intercal \mathbf{h}_{i,j}$, $\hat{\bm \phi}_3 = \mathbf{Q}^\intercal \bm \phi_3$.
The objective function is:
$$
J(\theta) = \frac{1}{2} \sum_i \|\hat{\mathbf{b}}_i(T) - \hat{\mathbf{d}}_i\|_2^2 + \lambda\sum_i \|\mathbf{w}_i\|_1.
$$
We denote
$$
\mathbf{x}_i(t) = \begin{bmatrix}\hat{\mathbf{b}}_i(t)\\ \hat{\mathbf{h}}_{i,1}(t)\\ \vdots\\ \hat{\mathbf{h}}_{i,s}(t)\end{bmatrix},
$$
and 
$$
\mathbf{M}_i(\theta) =
\begin{pmatrix}
\mathbf{w}_i^\intercal \bm{\kappa} \bm\Lambda + \mathbf{w}_i^\intercal \bm{\rho} \mathbf{I} & w_{i1} \mathbf{I} & w_{i2} \mathbf{I} & \cdots & w_{is} \mathbf{I} \\
\phi_1 \bm\Lambda & \bm\Lambda & \mathbf{0} & \cdots & \mathbf{0} \\
\phi_1 \bm\Lambda & \mathbf{0} & \bm\Lambda & \cdots & \mathbf{0} \\
\vdots & \vdots & \vdots & \ddots & \vdots \\
\phi_1 \bm\Lambda & \mathbf{0} & \mathbf{0} & \cdots & \bm\Lambda
\end{pmatrix},
$$
The original ODE can be transformed to
$$
\mathbf{\dot{x}}_i = \mathbf{M}_i(\theta)\mathbf{x}_i + \mathbf{g}_i,\quad \mathbf{x}_i(0)=\left[\hat{\mathbf{p}}_i, \mathbf{0}, \dots, \mathbf{0}\right]^\intercal,
$$
and $\mathbf{g}_i = [0, \hat{\bm\phi}_2, \dots, \hat{\bm\phi}_2]^\intercal$. The solution of the ODE at time $T$ is
\begin{equation*}
\mathbf{x}_i(T)
= e^{\mathbf{M}_i T}\,\mathbf{x}_{i0}
\;+\; \int_{0}^{T} e^{\mathbf{M}_i (T-\tau)}\,\mathbf{g}_i \,\mathrm d\tau.
\end{equation*}
Since $\mathbf{M}_i$ and $\mathbf{g}_i$ are constant in time, we have the following form
\begin{equation*}
\mathbf{x}_i(T)
= e^{\mathbf{M}_i T}\,\mathbf{x}_{i0} + e^{\mathbf{M}_iT}\int_{0}^{T} e^{-\mathbf{M}_i\tau}\,\mathrm d\tau \mathbf{g}_i,
\end{equation*}
and then we use composite trapezoidal rule to approximate the integral:
\begin{equation*}
\mathbf{x}_i(T)
= e^{\mathbf{M}_i T}\,\mathbf{x}_{i0} + e^{\mathbf{M}_iT}\sum_l \alpha_l e^{-\mathbf{M}_i\tau_l}\mathbf{g}_i.
\end{equation*}
We denote $\mathbf{Z}_T = e^{\mathbf{M}_i T}$ and $\bar{\mathbf{Z}}_T = \sum_l \alpha_l e^{-\mathbf{M}_i \tau_l}$.
Then the solution can be written as
\begin{equation*}
\mathbf{x}_i(T)
= \mathbf{Z}_T\,\mathbf{x}_{i0} + \mathbf{Z}_T\,\bar{\mathbf{Z}}_T\,\mathbf{g}_i.
\end{equation*}
We compute the derivative of the $\mathbf{Z}_T$ and $\bar{\mathbf{Z}}_T$ w.r.t all unknown parameters.
Let $\mathbf{P}_b\in\mathbb{R}^{N\times (s+1)N}$ pick the first $N$ entries, such that $\mathbf{P}_b\mathbf{x}_i(t)=\hat{\mathbf{b}}_i(t)$.
The residual can be reformulated as
$$
\mathbf{r}_i(\theta)=\mathbf{P}_b\,\mathbf{x}_i(T)-\hat{\mathbf{d}}_i,\qquad
$$
For any parameter \(\alpha\) and \(\beta\), the derivatives of the diagonal elements $\left[j, j\right]$ of \(Z_T\) and \(\bar{Z}_T\) are as follows:
\begin{align*}
\frac{\partial \mathbf{Z}_T}{\partial \alpha} &= T e^{\mathbf{M}_{i}T} \frac{\partial \mathbf{M}_{i}}{\partial \alpha}, \\
\frac{\partial \bar{\mathbf{Z}}_T}{\partial \alpha} &= -\sum_{\ell} \alpha_{\ell} \tau_{\ell} e^{-\mathbf{M}_{i} \tau_{\ell}}\frac{\partial \mathbf{M}_{i}}{\partial \alpha},
\end{align*}
The derivative of $\mathbf{x}_i(T)$ are as the following:
$$
\frac{\partial \mathbf{x}_i(T)}{\partial \alpha}
= \frac{\partial \mathbf{Z}_T}{\partial \alpha}\,\mathbf{x}_{i0}
+ \mathbf{Z}_T\,\frac{\partial \mathbf{x}_{i0}}{\partial \alpha}
+ \frac{\partial \mathbf{Z}_T}{\partial \alpha}\,\bar{\mathbf{Z}}_T\,\mathbf{g}_i
+ \mathbf{Z}_T\,\frac{\partial \bar{\mathbf{Z}}_T}{\partial \alpha}\,\mathbf{g}_i
+ \mathbf{Z}_T\,\bar{\mathbf{Z}}_T\,\frac{\partial \mathbf{g}_i}{\partial \alpha}.
$$
The output Jacobian is
$$
\mathbf{J}_{i,\alpha} = \mathbf{P}_b\,\frac{\partial \mathbf{x}_i(T)}{\partial \alpha}.
$$
Specifically, we write down the non-zero terms of the Jacobian of each model parameter.
Jacobian w.r.t.\ initial condition $p_i$ is
\begin{align*}
\frac{\partial \mathbf{x}_i(T)}{\partial p_i} &= \mathbf{Z}_T \mathbf{P}_b^\intercal, \\
J_{i,p_i} &= \mathbf{P}_b \mathbf{Z}_T \mathbf{P}_b^\intercal.
\end{align*}
Jacobian w.r.t.\ $w_{ij}$ is
\begin{align*}
\frac{\partial \mathbf{M}_i}{\partial w_{ij}} &=
\begin{pmatrix}
\kappa_j \bm\Lambda + \rho_j \mathbf{I} & \mathbf{0} & \cdots & \mathbf{0} & \mathbf{I}_{(j)} & \mathbf{0} & \cdots \mathbf{0} \\
\mathbf{0} & \mathbf{0} & \cdots & \mathbf{0} & \mathbf{0} & \mathbf{0} & \cdots \mathbf{0} \\
\vdots & \vdots & \cdots & \vdots & \vdots & \vdots & \cdots \vdots\\
\mathbf{0} & \mathbf{0} & \cdots & \mathbf{0} & \mathbf{0} & \mathbf{0} & \cdots \mathbf{0}
\end{pmatrix}, \\
\frac{\partial \mathbf{x}_i(T)}{\partial w_{ij}} &=
\frac{\partial \mathbf{Z}_T}{\partial w_{ij}}(\mathbf{x}_{i0} + \bar{\mathbf{Z}}_T \mathbf{g}_i) + \mathbf{Z}_T \frac{\partial \bar{\mathbf{Z}}_T}{\partial w_{ij}} \mathbf{g}_i, \\
\mathbf{J}_{i,w_{ij}} &= \mathbf{P}_b\,\frac{\partial \mathbf{x}_i(T)}{\partial w_{ij}}.
\end{align*}
Jacobian w.r.t.\ diffusion coefficients $\kappa_j$ is
\begin{align*}
\frac{\partial \mathbf{M}_i}{\partial \kappa_j} &=
\begin{pmatrix}
w_{ij}\bm\Lambda & \mathbf{0} & \cdots & \mathbf{0}\\
\mathbf{0} & \mathbf{0} & \cdots & \mathbf{0}\\
\vdots & \vdots & \ddots & \vdots\\
\mathbf{0} & \mathbf{0} & \cdots & \mathbf{0}
\end{pmatrix}, \\
\frac{\partial \mathbf{x}_i(T)}{\partial \kappa_j} &= \frac{\partial \mathbf{Z}_T}{\partial \kappa_j}(\mathbf{x}_{i0}+\bar{\mathbf{Z}}_T \mathbf{g}_i), \\
\mathbf{J}_{i,\kappa_j} &= \mathbf{P}_b\,\frac{\partial \mathbf{x}_i(T)}{\partial \kappa_j}.
\end{align*}
Jacobian w.r.t.\ growth rates $\rho_j$ is
\begin{align*}
\frac{\partial \mathbf{M}_i}{\partial \rho_j} &=
\begin{pmatrix}
w_{ij} \mathbf{I} & \mathbf{0} & \cdots & \mathbf{0}\\
\mathbf{0} & \mathbf{0} & \cdots & \mathbf{0}\\
\vdots & \vdots & \ddots & \vdots\\
\mathbf{0} & \mathbf{0} & \cdots & \mathbf{0}
\end{pmatrix}, \\
\frac{\partial \mathbf{x}_i(T)}{\partial \rho_j} &= \frac{\partial \mathbf{Z}_T}{\partial \rho_j}(\mathbf{x}_{i0}+\bar{\mathbf{Z}}_T \mathbf{g}_i), \\
\mathbf{J}_{i,\rho_j} &= \mathbf{P}_b\,\frac{\partial \mathbf{x}_i(T)}{\partial \rho_j}.
\end{align*}
Jacobian w.r.t.\ coupling $\phi_1$ is
\begin{align*}
\frac{\partial \mathbf{M}_i}{\partial \phi_1} &=
\begin{pmatrix}
\mathbf{0} & \mathbf{0} & \cdots & \mathbf{0}\\
\bm\Lambda & \mathbf{0} & \cdots & \mathbf{0}\\
\vdots & \vdots & \ddots & \vdots\\
\bm\Lambda & \mathbf{0} & \cdots & \mathbf{0}
\end{pmatrix}, \\
\frac{\partial \mathbf{x}_i(T)}{\partial \phi_1} &= \frac{\partial \mathbf{Z}_T}{\partial \phi_1}(\mathbf{x}_{i0}+\bar{\mathbf{Z}}_T \mathbf{g}_i), \\
\mathbf{J}_{i,\phi_1} &= \mathbf{P}_b\,\frac{\partial \mathbf{x}_i(T)}{\partial \phi_1}.
\end{align*}
Jacobian w.r.t.\ bias term $\phi_3$ is
\begin{align*}
\frac{\partial \mathbf{g}_i}{\partial \phi_3} &= \begin{bmatrix}\mathbf{0} \\ \mathbf{I} \\ \vdots \\ \mathbf{I}\end{bmatrix}, \\
\frac{\partial \mathbf{x}_i(T)}{\partial \phi_3} &= \mathbf{Z}_T \bar{\mathbf{Z}}_T \frac{\partial \mathbf{g}_i}{\partial \phi_3}, \\
\mathbf{J}_{i,\phi_3} &= \mathbf{P}_b \mathbf{Z}_T \bar{\mathbf{Z}}_T \frac{\partial \mathbf{g}_i}{\partial \phi_3}.
\end{align*}

The Hessian matrix of the objective function with respect to the parameters $\bm\theta$ is given by
$$
\mathbf{H} =
\left[
\begin{array}{ccccccccccc}
\mathbf{H}_{\mathbf{\hat{p}}_1\mathbf{\hat{p}}_1} & \mathbf{H}_{\mathbf{\hat{p}}_1\mathbf{w}_1} & \mathbf{0} & \cdots & \mathbf{0} & \mathbf{0} & \mathbf{0} & \mathbf{H}_{\mathbf{\hat{p}}_1\bm\kappa} & \mathbf{H}_{\mathbf{\hat{p}}_1\bm\rho} & \mathbf{H}_{\mathbf{\hat{p}}_1\bm\phi_1} & \mathbf{H}_{\mathbf{\hat{p}}_1\bm\phi_3} \\
\mathbf{H}_{\mathbf{w}_1\mathbf{\hat{p}}_1} & \mathbf{H}_{\mathbf{w}_1\mathbf{w}_1} & \mathbf{0} & \cdots & \mathbf{0} & \mathbf{0} & \mathbf{0} & \mathbf{H}_{\mathbf{w}_1\bm\kappa} & \mathbf{H}_{\mathbf{w}_1\bm\rho} & \mathbf{H}_{\mathbf{w}_1\bm\phi_1} & \mathbf{H}_{\mathbf{w}_1\bm\phi_3} \\
\mathbf{0} & \mathbf{0} & \mathbf{H}_{\mathbf{\hat{p}}_2\mathbf{\hat{p}}_2} & \mathbf{H}_{\mathbf{\hat{p}}_2\mathbf{w}_2} & \cdots & \mathbf{0} & \mathbf{0} & \mathbf{H}_{\mathbf{\hat{p}}_2\bm\kappa} & \mathbf{H}_{\mathbf{\hat{p}}_2\bm\rho} & \mathbf{H}_{\mathbf{\hat{p}}_2\bm\phi_1} & \mathbf{H}_{\mathbf{\hat{p}}_2\bm\phi_3} \\
\mathbf{0} & \mathbf{0} & \mathbf{H}_{\mathbf{w}_2\mathbf{\hat{p}}_2} & \mathbf{H}_{\mathbf{w}_2\mathbf{w}_2} & \cdots & \mathbf{0} & \mathbf{0} & \mathbf{H}_{\mathbf{w}_2\bm\kappa} & \mathbf{H}_{\mathbf{w}_2\bm\rho} & \mathbf{H}_{\mathbf{w}_2\bm\phi_1} & \mathbf{H}_{\mathbf{w}_2\bm\phi_3} \\
\vdots & \vdots & \vdots & \vdots & \ddots & \vdots & \vdots & \vdots & \vdots & \vdots & \vdots \\
\mathbf{0} & \mathbf{0} & \mathbf{0} & \mathbf{0} & \cdots & \mathbf{H}_{\mathbf{\hat{p}}_n\mathbf{\hat{p}}_n} & \mathbf{H}_{\mathbf{\hat{p}}_n\mathbf{w}_n} & \mathbf{H}_{\mathbf{\hat{p}}_n\bm\kappa} & \mathbf{H}_{\mathbf{\hat{p}}_n\bm\rho} & \mathbf{H}_{\mathbf{\hat{p}}_n\bm\phi_1} & \mathbf{H}_{\mathbf{\hat{p}}_n\bm\phi_3} \\
\mathbf{0} & \mathbf{0} & \mathbf{0} & \mathbf{0} & \cdots & \mathbf{H}_{\mathbf{w}_n\mathbf{\hat{p}}_n} & \mathbf{H}_{\mathbf{w}_n\mathbf{w}_n} & \mathbf{H}_{\mathbf{w}_n\bm\kappa} & \mathbf{H}_{\mathbf{w}_n\bm\rho} & \mathbf{H}_{\mathbf{w}_n\bm\phi_1} & \mathbf{H}_{\mathbf{w}_n\bm\phi_3} \\
\mathbf{H}_{\bm\kappa\mathbf{\hat{p}}_1} & \mathbf{H}_{\bm\kappa\mathbf{w}_1} & \mathbf{H}_{\bm\kappa\mathbf{\hat{p}}_2} & \mathbf{H}_{\bm\kappa\mathbf{w}_2} & \cdots & \mathbf{H}_{\bm\kappa\mathbf{\hat{p}}_n} & \mathbf{H}_{\bm\kappa\mathbf{w}_n} & \mathbf{H}_{\bm\kappa\bm\kappa} & \mathbf{H}_{\bm\kappa\bm\rho} & \mathbf{H}_{\bm\kappa\bm\phi_1} & \mathbf{H}_{\bm\kappa\bm\phi_3} \\
\mathbf{H}_{\bm\rho\mathbf{\hat{p}}_1} & \mathbf{H}_{\bm\rho\mathbf{w}_1} & \mathbf{H}_{\bm\rho\mathbf{\hat{p}}_2} & \mathbf{H}_{\bm\rho\mathbf{w}_2} & \cdots & \mathbf{H}_{\bm\rho\mathbf{\hat{p}}_n} & \mathbf{H}_{\bm\rho\mathbf{w}_n} & \mathbf{H}_{\bm\rho\bm\kappa} & \mathbf{H}_{\bm\rho\bm\rho} & \mathbf{H}_{\bm\rho\bm\phi_1} & \mathbf{H}_{\bm\rho\bm\phi_3} \\
\mathbf{H}_{\bm\phi_1\mathbf{\hat{p}}_1} & \mathbf{H}_{\bm\phi_1\mathbf{w}_1} & \mathbf{H}_{\bm\phi_1\mathbf{\hat{p}}_2} & \mathbf{H}_{\bm\phi_1\mathbf{w}_2} & \cdots & \mathbf{H}_{\bm\phi_1\mathbf{\hat{p}}_n} & \mathbf{H}_{\bm\phi_1\mathbf{w}_n} & \mathbf{H}_{\bm\phi_1\bm\kappa} & \mathbf{H}_{\bm\phi_1\bm\rho} & \mathbf{H}_{\bm\phi_1\bm\phi_1} & \mathbf{H}_{\bm\phi_1\bm\phi_3} \\
\mathbf{H}_{\bm\phi_3\mathbf{\hat{p}}_1} & \mathbf{H}_{\bm\phi_3\mathbf{w}_1} & \mathbf{H}_{\bm\phi_3\mathbf{\hat{p}}_2} & \mathbf{H}_{\bm\phi_3\mathbf{w}_2} & \cdots & \mathbf{H}_{\bm\phi_3\mathbf{\hat{p}}_n} & \mathbf{H}_{\bm\phi_3\mathbf{w}_n} & \mathbf{H}_{\bm\phi_3\bm\kappa} & \mathbf{H}_{\bm\phi_3\bm\rho} & \mathbf{H}_{\bm\phi_3\bm\phi_1} & \mathbf{H}_{\bm\phi_3\bm\phi_3}
\end{array}
\right].
$$
For any parameters $\alpha,\beta$,
$$
\mathbf{H}_{\alpha\beta} = \sum_{i=1}^n 
\left[
\mathbf{J}_{i,\alpha}^T \mathbf{J}_{i,\beta}
+ \mathbf{r}_i^T \mathbf{P}_b \frac{\partial^2 \mathbf{x}_i(T)}{\partial \beta \partial \alpha}
\right].
$$
We assume the system is noise-free and therefore $\mathbf{r}_i=\mathbf{0}$.
We compute the numerical condition number of the Hessian matrix \( \mathbf{H} \) to assess the invertibility of the optimization problem~\citep{lehoucq1998arpack}.
A large condition number indicates that the Hessian is close to singular, which can lead to numerical instability in optimization.
On the other hand, a moderate condition number suggests that the optimization problem is well-posed and that the parameters can be reliably estimated from the data.

\subsection{Numerical Discretization and Inversion Solver} 
Time discretization of the forward ODEs is performed using the LSODA solver~\cite{petzold1983automatic}. 
Gradients are computed with JAX~\cite{jax2018github}, and optimization is carried out using the limited-memory quasi-Newton L-BFGS method~\cite{zhu1997algorithm}. 
The $\ell_0$ constraints $\|\mathbf{p}_i\|_0=s^{\text{max}}$ and $\|\mathbf{w}_{h,i}\|_0=1$ are enforced via compressed sampling techniques~\cite{tropp2010computational}.
More details about the compressed sampling algorithm on pathological spreading are extensively discussed in~\cite{wen2025single}.
We test different choices of the number of latent states \( m = 0, 1, 3, 5, 10 \) and for two brain atlases: MUSE and DKT.

\subsection{Code parallelization.}
From a computational complexity perspective, because the inversion algorithm leverages cohort-level data to update model parameters, the computational cost increases with the number of subjects. 
To address this challenge, our implementation supports parallel computing via the Message Passing Interface (MPI). 
Since the gradient computation for each subject is independent, we distribute the gradient computation for different subjects to different CPU cores.
During each optimization iteration, we gather all gradients to the master core to update the parameters.
In particular, the number of MPI processes can be set to match the number of subjects, enabling subject-level parallelism.
For example, inverting data from 500 clinical subjects with \( m = 1 \) latent state requires more than 12 hours when using a sequential implementation. 
In contrast, when executed with 500 MPI processes, the same inversion task completes in less than 10 minutes, demonstrating the scalability and computational efficiency of the parallelized algorithm.

\section{Results}
We first verify the proposed method on synthetic data regarding the accuracy of inversion algorithm, parameter identifiability and model stability.
We then evaluated the proposed model on two clinical datasets including ADNI and A4 study. 
Especially, we test the model forecast ability for unseen scan in the clinical dataset.
We in additionally reveal the common initial seed of the disease. 
From the latent state analysis, we find subtypes of the disease trajectories.

\begin{figure}[!htbp]
\centering
\includegraphics[width=1\textwidth]{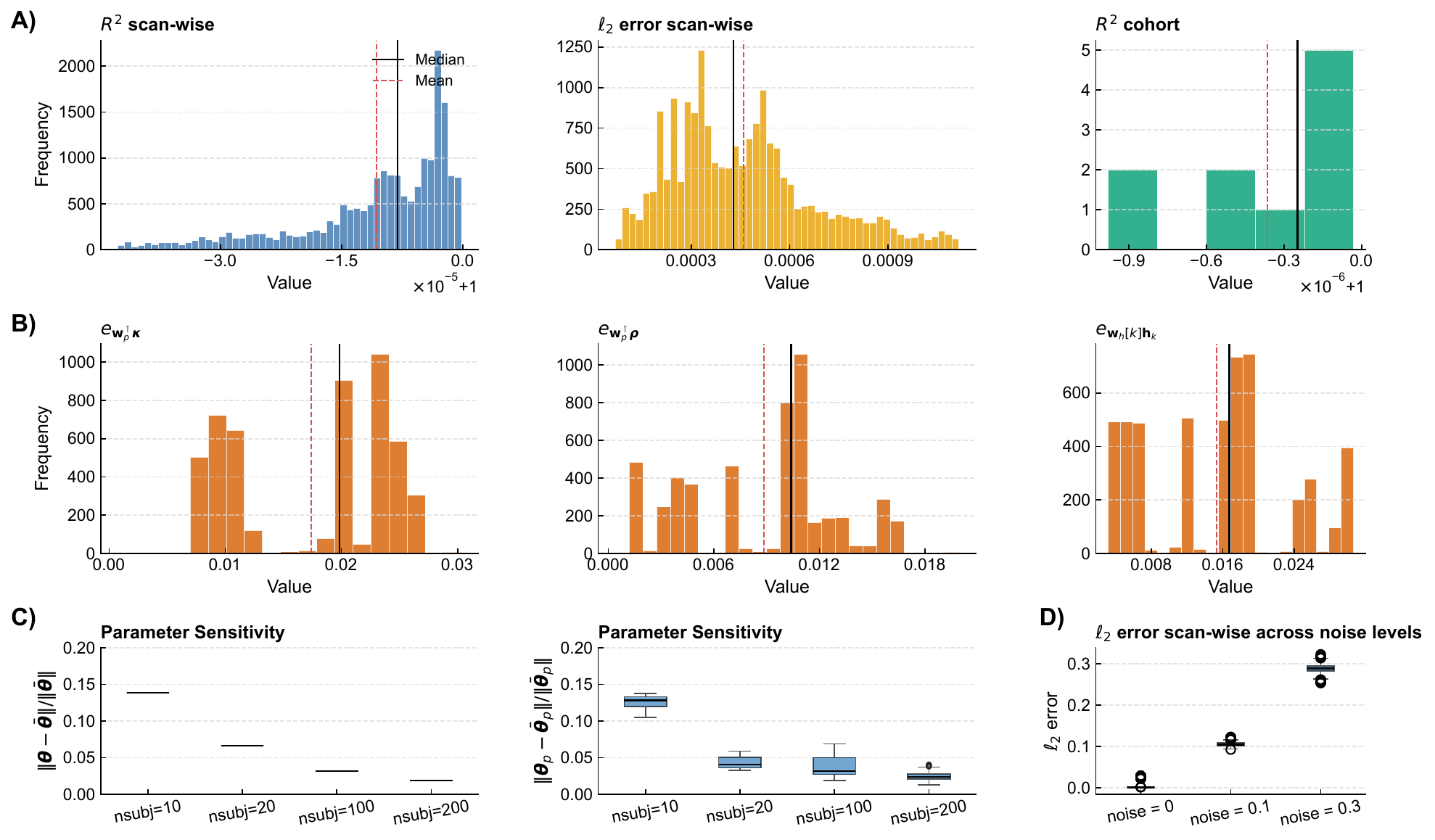}
\caption{
  \textbf{Synthetic experiment results.}
  We generate synthetic data using the proposed LNODE model with \( m = 1 \) latent state and randomly sampled, known cohort-shared and subject-specific parameters. 
  We then apply the LNODE inversion procedure to recover the synthetic scans and model parameters. 
  Model performance is evaluated in terms of reconstruction accuracy and stability of the estimated parameters. 
  With fixed observation data, we repeat the inversion procedure ten times using different random initializations.
  \textbf{(A)} From left to right, we assess scan reconstruction accuracy using histograms of the per-scan \( R_{\text{scan}}^2 \), the scan-wise relative error \( e_{\text{scan}} \), and the cohort-level \( R_{\text{cohort}}^2 \).
  \textbf{(B)} From left to right, we evaluate parameter estimation accuracy using histograms of the relative parameter error \( e_{\bTh} \) for the estimated quantities \( \mathbf{w}_p^\intercal \boldsymbol{\kappa} \), \( \mathbf{w}_p^\intercal \boldsymbol{\rho} \), and \( \mathbf{w}_h[k]\,\mathbf{h}_k \).
  \textbf{(C)} From left to right, we show the stability of the estimated parameters \( \bTh \) and \( \bTh_p \) across different random initializations, as well as the scan reconstruction error under additive Gaussian noise with varying standard deviations.
}
\label{fig:synthetic_results}
\end{figure}

\begin{figure}[!htbp]
  \centering
  \includegraphics[width=\columnwidth]{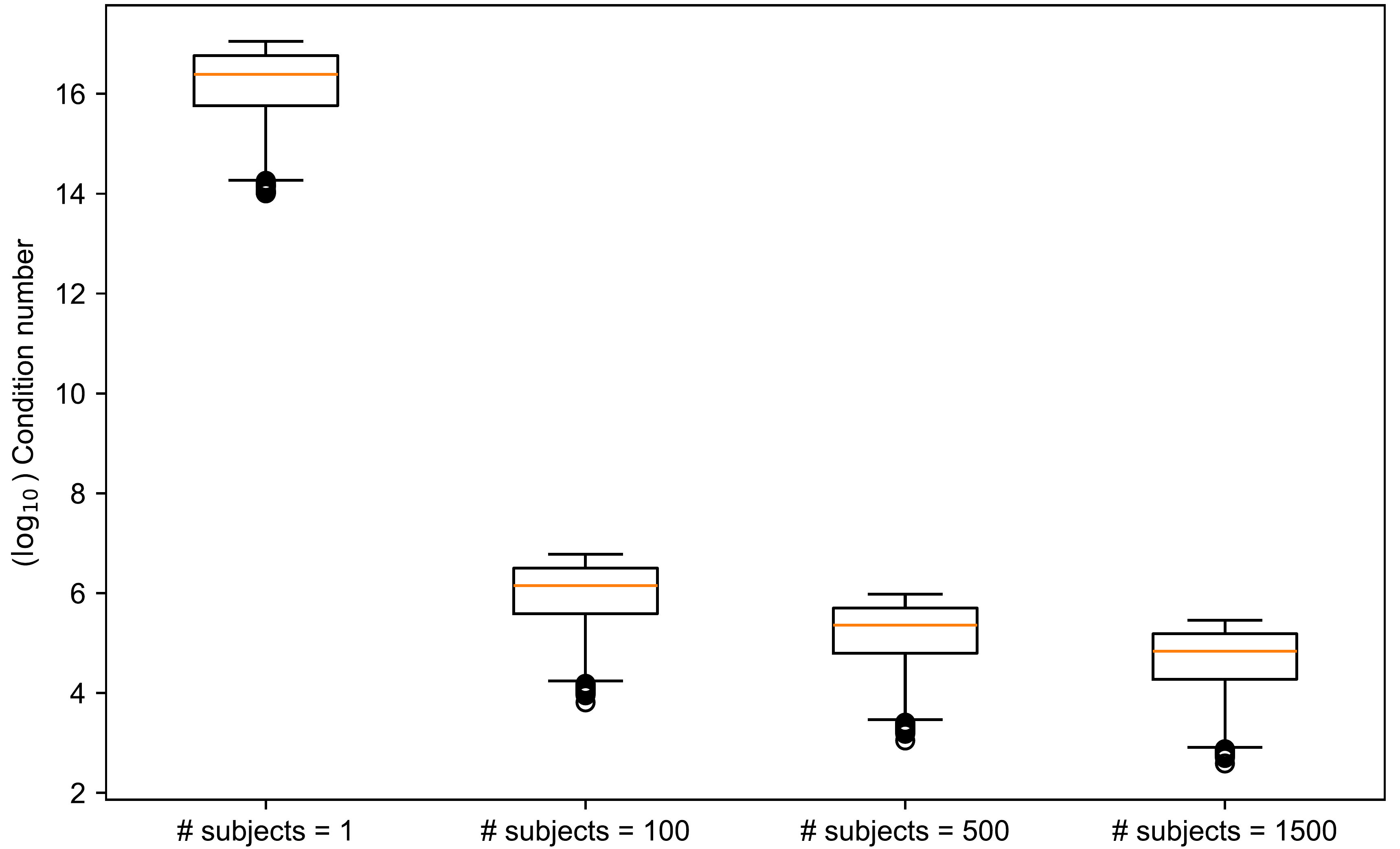}
  \caption{We compute the condition number of the whole Hessian matrix $\mathbf{H}$. 
  We explore the changes of the condition number versus number of subjects.
  The condition number decreses as the number of subjects increases, indicating improved identifiability and stability of parameter estimation.
  }
  \label{fig:invertibility_hessian}
\end{figure}

\subsection{LNODE estimates accurate model parameters.}
We validate the accuracy of the cohort inversion algorithm through synthetic experiments. 
The synthetic data are generated from equation~(\ref{eq:forward}) with randomly sampled ground-truth parameters to simulate realistic PET observations. 
We generate 500 synthetic subjects and each subject has four scans. 
In this setting, we assume that there is no model error.
We apply the inversion algorithm to recover the synthetic scans and the corresponding model parameters. 
We evaluate the inversion algorithm from two perspectives: (1) reconstruction accuracy of both parameters and observation data, and (2) stability of the estimated parameters under different initializations.

The scan reconstruction accuracy is reported in Fig.~\ref{fig:synthetic_results}(A). 
Both the per-scan \( R_{\text{scan}}^2 \) and cohort-level \( R_{\text{cohort}}^2 \) exceed 0.9999 (with 1 being the maximum possible value). 
The average relative \( \ell_2 \)-norm reconstruction error, \( e_{\text{scan}} \), is \( 4 \times 10^{-4} \) across all scans (with 0 being the minimum possible value).
The parameter estimation accuracy is shown in Fig.~\ref{fig:synthetic_results}(B). 
For key quantities, including \( \mathbf{w}_p^\intercal \boldsymbol{\kappa} \), \( \mathbf{w}_p^\intercal \boldsymbol{\rho} \), and \( \mathbf{w}_h[k]\mathbf{h}_k \), the average relative parameter error \( e_{\boldsymbol{\theta}} \) remains below 0.02.
To assess robustness, we further add Gaussian noise with varying standard deviations to the synthetic scans and repeat the inversion procedure. 
As shown in the last column of Fig.~\ref{fig:synthetic_results}(C), the reconstruction error scales consistently with the noise level, demonstrating the robustness of the inversion algorithm. 
In addition, we evaluate the stability of the estimated parameters across different random initializations. 
We generate synthetic datasets with \( n = 10, 20, 100, 200 \) subjects, each again with four observation scans. 
Under a noise-free setting, we quantify the stability of the estimated parameters across different random initializations using the metrics \( \Xi_{\mathbf{\Theta}} \) and \( \Xi_{\mathbf{\Theta}_p} \).
We observe that increasing the number of subjects included in the cohort inversion leads to improved parameter stability.
Specifically, we compute the relative difference between the estimated parameters obtained from each random initialization and the mean estimate across all initializations. 
The results in the first two columns of Fig.~\ref{fig:synthetic_results}(C) indicate that the instability of the estimated cohort-shared parameters \( \bTh \) decreases from 0.14 to 0.02 as the number of subjects increases from 10 to 200. 
Similarly, the instability of the subject-specific parameters \( \bTh_p \) decreases from 0.13 to 0.03.

We further evaluate the identifiability of the estimated parameters by the condition number of the analytical hessian as detailed in Section~\ref{subsec:invertibility}.
We present the results of the condition number analysis in Fig. \ref{fig:invertibility_hessian}.
As we increase the number of subjects in the cohort inversion, the condition number of the Hessian matrix decreases from the level of $10^{16}$ to $10^4$ as we increase the number of subjects from $1$ to $1500$, indicating improved identifiability of the parameters.

\subsection{LNODE successfully models \Ab PET dynamics.}
We examine the model performance on clinical datasets: ADNI and A4 study.
From the ADNI cohort, we selected 1,461 subjects who have T1-weighted MRI and \Ab PET scans.
It includes 548 subjects with cognitively normal (CN), 737 subjects with mild cognitive impairment (MCI), and 176 subjects with Alzheimer's disease (AD). 
The mean ages of the CN, MCI, and AD groups are 73.93, 73.44, and 74.97 years, respectively. 
The mean scores for the Mini-Mental State Examination (MMSE) are 28.84, 27.17, and 21.66.
We include 1070 study subjects from both A4 whose MRI and \Ab PET scans are available. 
They were cognitively normal at the time of enrollment.
The mean age of the A4 cohort is 71.92 years, with a mean MMSE score of 28.58. 
Additional demographic details for both cohorts are summarized in Table~\ref{tab:demograph}.

\begin{table*}[!htbp]
  \caption{
    \textbf{Demographic summary of ADNI and A4 datasets with \Ab PET scans.}
      We collect 1461 subjects from the ADNI dataset. 
      These subjects have both T1 MRI and \Ab PET scans. 
      In the table, we summarize demographic information of age, sex, Mini-Mental State Examination (MMSE), 
      Alzheimer's Disease Assessment Scale 13 (ADAS13), ADNI Execution Function (ADNI EF) and ADNI Memory score (ADNI Mem).
      We also collect 1070 subjects from the A4 Study dataset. 
      We summarize demographic information of age, sex and MMSE for A4 subjects in the table.
  }
  \begin{adjustbox}{width=0.8\textwidth,center}
  \small \setlength{\tabcolsep}{6pt}
  \begin{tabular}{c|c|c|c}
  \toprule
  \multicolumn{4}{c}{\textbf{ADNI Dataset ($n=1461$)}} \\
  \midrule
  Characteristic & CN ($n=548$) & MCI ($n=737$) & AD ($n=176$) \\
  \midrule 
  Age (years) & 73.93 ($\pm$ 7.78) & 73.44 ($\pm$ 7.58) & 74.97 ($\pm$ 8.47) \\
  Sex (male / female) & 224 / 324 & 398 / 339 & 100 / 76 \\
  MMSE & 28.84 ($\pm$ 1.10) & 27.17 ($\pm$ 2.48) & 21.66 ($\pm$ 3.08) \\
  ADAS13 & 9.81 ($\pm$ 5.55) & 16.85 ($\pm$ 9.89) & 33.61 ($\pm$ 9.39) \\
  ADNI EF & 0.76 ($\pm$ 0.89) & 0.28 ($\pm$ 0.89) & $-1.04$ ($\pm$ 0.92) \\
  ADNI Mem & 0.90 ($\pm$ 0.77) & 0.31 ($\pm$ 0.85) & $-0.95$ ($\pm$ 0.59) \\
  \midrule
  \multicolumn{4}{c}{\textbf{A4 Dataset ($n=1070$)}} \\
  \midrule
  Age (years) & \multicolumn{3}{c}{71.92 ($\pm$ 4.74)} \\
  Sex (male / female) & \multicolumn{3}{c}{434 / 636} \\
  MMSE & \multicolumn{3}{c}{28.58 ($\pm$ 1.93)} \\
  \bottomrule
  \end{tabular}
  \end{adjustbox}
  \label{tab:demograph}
\end{table*}

\begin{figure}[!htbp]
\centering
\includegraphics[width=1\textwidth]{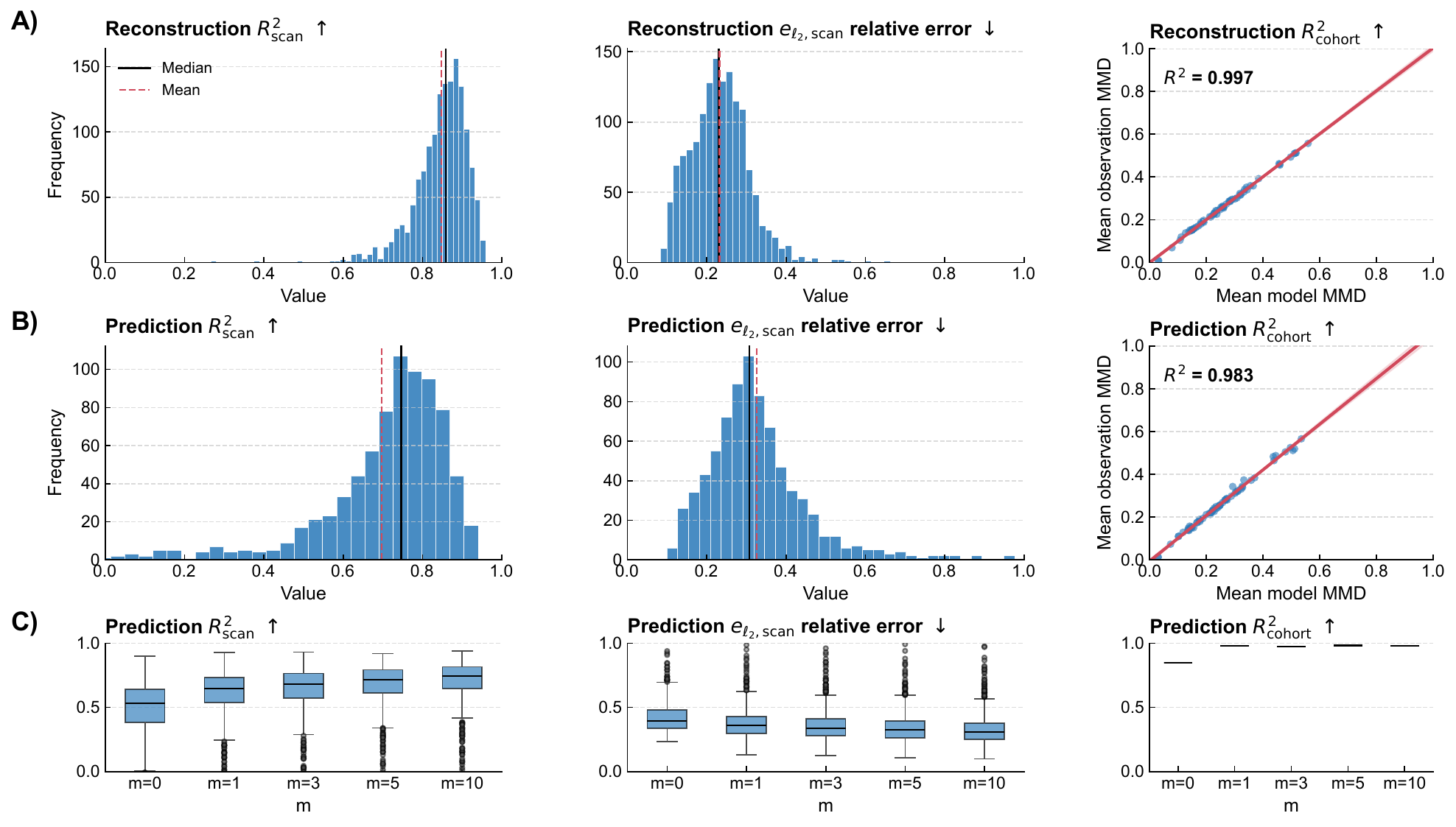}
\caption{
  \textbf{Quantitative evaluation of the model reconstruction and prediction performance using the DKT atlas.} 
  We conduct two experiments to evaluate the model performance on the ADNI cohort: In the first one we reconstruct the data using all cohort scans; in the second one, we use LNODE to predict the second scan using only the first scan to estimate model parameters (defined as the {\bf `Pred1'} task in the text). 
  The top row shows the reconstruction performance with ten latent states (\( m = 10 \)), while the middle row presents the prediction performance for the next scan with \( m = 10 \). 
  The bottom row shows the prediction performance for the next scan for different numbers of latent states, \( m = 0, 1, 3, 5, 10 \).
  From left to right, the panels display the distribution of per-scan \( R_{\text{scan}}^2 \) scores (higher is better), the distribution of per-scan relative errors \( e_{\text{scan}} \) (lower is better), and the cohort-level \( R_{\text{cohort}}^2 \) score (higher is  better). 
  Overall, the model achieves excellent reconstruction and prediction performance, characterized by high \( R^2 \) scores and low relative errors. 
  Moreover, increasing the number of latent states \( m \) consistently improves prediction performance, indicating that a larger latent space enhances the model's ability to capture complex \Ab dynamics.
}
\label{fig:clinical_quantitative_inversion_results}
\end{figure}

In this study, when we mention `Reconstruction', we refer to the model's ability to fit all available scans from each subject within the cohort.
`Prediction' refers to the model's ability to forecast unseen future scans for each subject.
In the `Prediction' task, we evaluate three scenarios: 
\begin{itemize}
\item {\bf (Pred1)}: First we train LNODE using only the \emph{first scan} from all subjects in the cohort.
We estimate both the cohort-shared parameters $\bTh_c$ and the subject-specific parameters $\bTh_{p,i}$. 
  Using the trained LNODE for each subject  we predict the \emph{second scan}.

\item {\bf (Pred2)}: We train LNODE using only the \emph{first scan} of all subjects in the cohort.
  In this phase, we estimate both the parameters shared by the cohort $\bTh_c$ and the subject-specific parameters $\bTh_{p,i}$. 
  Using the trained LNODE, for each subject, we predict the \emph{last scan}. This run we discuss only in the text since it largely overlaps with Pred1. 
  For most subjects, we only have two scans. On average, ADNI subjects have 2.19 scans per individual.

\item {\bf (Pred3)}: We split the cohort into two subsets: training and testing.
  We use the training set to estimate $\bTh_c$ and subject-specific parameters $\bTh_{p,i}$.
  Then for the subjects in the testing set, we fix the already estimated $\bTh_c$ and calibrate $\bTh_{p,i}$ using \emph{only the first scan}. Then we predict the \emph{next scan}. 

\end{itemize}
We will refer to the three prediction tasks as the prediction tasks `Pred1', `Pred2', and `Pred3', respectively.
We use several metrics to evaluate the performance of the model. 
Specifically, $R_{\text{scan}}^2$ and $R_{\text{cohort}}^2$ denote the coefficient of determination at the scan and cohort levels, respectively (higher values indicate better performance). 
The metric $e_{\text{scan}}$ measures the relative error for each scan (lower values are better), while $e_{\boldsymbol{\theta}}$ quantifies the relative error of the estimated model parameters (lower values indicate more accurate parameter recovery).
\par
We quantitatively evaluated the performance of the reconstruction and prediction (`Pred1' task) of the proposed model in the ADNI cohort.
Fig.~\ref{fig:clinical_quantitative_inversion_results} summarizes the quantitative results obtained using the DKT atlas. 
The performance of the model is assessed using three metrics: the per-scan \( R_{\text{scan}}^2 \), the relative error per-scan \( e_{\text{scan}} \), and the cohort-level \( R_{\text{cohort}}^2 \).
The reconstruction experiment (top row of Fig.~\ref{fig:clinical_quantitative_inversion_results}) uses ten latent states (\( m = 10 \)). 
The mean, median, and standard deviation of \( R_{\text{scan}}^2 \) are 0.847, 0.859, and 0.066, respectively. 
The corresponding statistics for the relative error \( e_{\text{scan}} \) are 0.232, 0.231, 0.072. 
Performance at the cohort-level is high, with \( R_{\text{cohort}}^2 = 0.997 \).
\par
In the prediction experiment (second row of Fig.~\ref{fig:clinical_quantitative_inversion_results}, `Pred1' task), the mean, median, and standard deviation of \( R_{\text{scan}}^2 \) are 0.697, 0.746, and 0.193, respectively. 
The mean \( e_{\text{scan}} \) is 0.326, with a median of 0.307 and a standard deviation of 0.124. 
The cohort-level prediction accuracy of the next scan remains strong, with \( R_{\text{cohort}}^2 = 0.983 \).
Additionally, in the prediction task for the last scan of each subject (`Pred2' task), we have \( R_{\text{cohort}}^2 = 0.939 \) for the DKT atlas.
\par
We further evaluate the prediction performance (for `Pred1`) under different numbers of latent states, \( m = 0, 1, 3, 5, 10 \) (bottom row of Fig.~\ref{fig:clinical_quantitative_inversion_results}). 
As the number of latent states increases, prediction performance consistently improves across all three metrics. 
In the absence of latent states (\( m = 0 \)), the mean, median, and standard deviation of \( R_{\text{scan}}^2 \) are 0.485, 0.532, and 0.229, respectively. 
Increasing the number of latent states to \( m = 1 \) improves the mean \( R_{\text{scan}}^2 \) to 0.602, the median to 0.646, and reduces the standard deviation to 0.198.
At the cohort level, \( R_{\text{cohort}}^2 \) increases substantially from 0.848 for \( m = 0 \) to 0.980 for \( m = 1 \).
In particular, this performance gain is achieved by adding only one additional subject-specific parameter per subject. 
The number of parameters shared by the cohort increases by \( N + 2 = 64 \), which is negligible in terms of the complexity of the model relative to the size of the cohort. 
Prediction performance continues to improve as the number of latent states increases to \( m = 3, 5, 10 \).
However, further increasing the number of latent states may introduce overfitting. 
In addition, the performance gain of the model becomes less pronounced as \( m \) increases.
We also evaluate the quantitative performance of the proposed model using the MUSE atlas, with results reported in the Supplementary Fig.~\ref{fig:inversion_results_muse}. 
The performance of the proposed model is consistent with that obtained using the DKT atlas.
Notably, because the MUSE atlas contains a larger number of ROIs than the DKT atlas (98 vs 62), the benefit of incorporating latent states is more pronounced.
\par
\begin{figure}[!htbp]
\centering
\includegraphics[width=1\textwidth]{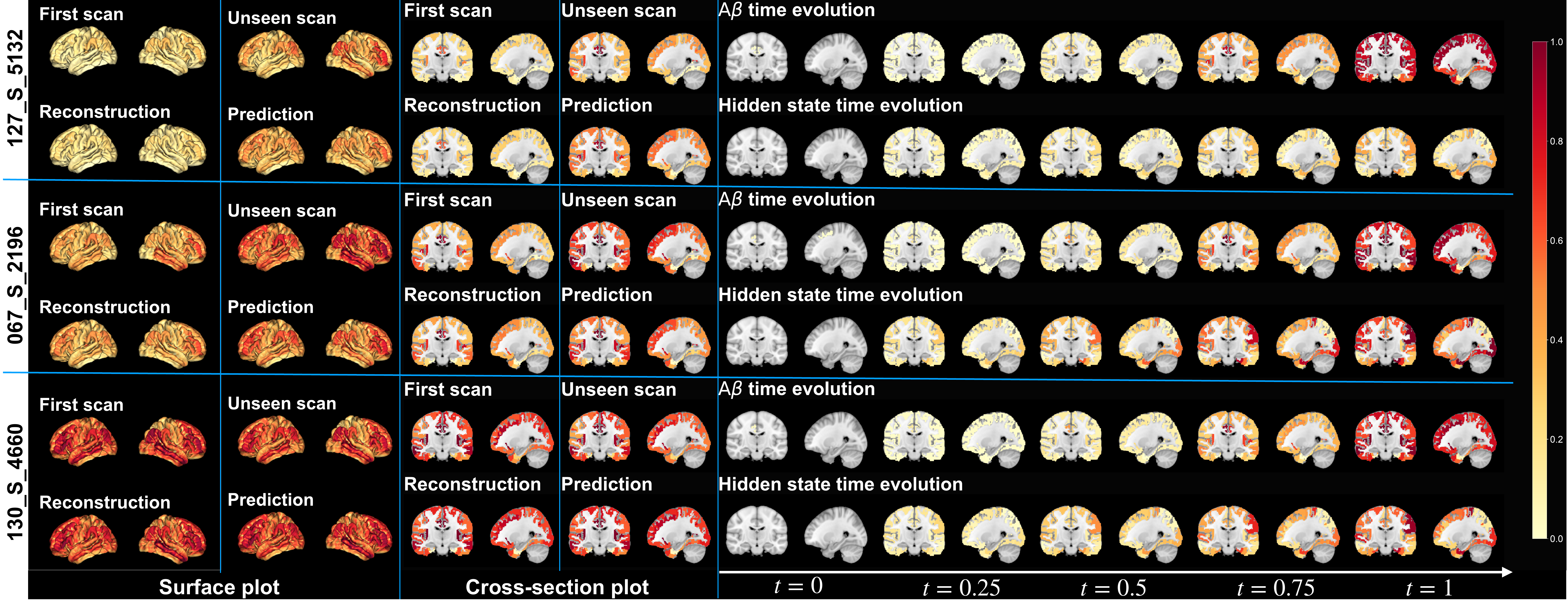}
\caption{
  \textbf{Qualitative evaluation of the model reconstruction and prediction performance using the DKT atlas.} 
  We perform model inversion using the first scan of each subject as input and predict the next scan ({\bf `Pred1'} task) for subjects in the ADNI cohort. 
  From top to bottom, the three rows correspond to three representative subjects (127\_S\_5132, 067\_S\_2169, and 130\_S\_4660), selected from the CN, MCI, and AD groups, respectively. 
  These subjects exhibit substantial changes in \Ab burden between the first and next scans.
  From left to right, the figure contains five panels: surface plots of reconstruction performance, surface plots of prediction performance, cross-section plots of reconstruction performance, cross-section plots of prediction performance, and the estimated trajectories of \Ab burden and latent states across disease age \( t \). 
  Each of the first four panels spans two columns to display two different brain views. 
  The final panel spans ten columns, illustrating the estimated abnormal \Ab trajectory \( \mathbf{b}_a(t) \) and the selected latent state trajectory \( \sum_k \mathbf{w}_{h}[k]\mathbf{h}_k(t) \) at five disease-age time points \( t = 0, 0.25, 0.5, 0.75, 1 \), with two brain views shown at each time point.
  Overall, the model successfully reconstructs and predicts \Ab burden patterns for individual subjects. 
  The estimated trajectory plots further provide insight into the interaction between the latent states and subject-specific \Ab progression.
}
\label{fig:clinical_visualization_pred_dkt}
\end{figure}

Figure~\ref{fig:clinical_visualization_pred_dkt} (first vertical panel denoted by \emph{First scan} and \emph{Reconstruction}) presents qualitative reconstruction results for three representative subjects of the CN, MCI, and AD groups using the DKT atlas. 
These subjects are selected because they exhibit substantial changes in \Ab burden between the first and next scans. 
In general, the model successfully reconstructs \Ab burden patterns at the individual-subject level.
Using the estimated model parameters, we further predict future unseen \Ab scans for each subject (`Pred1' task). 
The corresponding prediction results are shown in the second vertical panel of Fig.~\ref{fig:clinical_visualization_pred_dkt} (denoted by  \emph{Unseen scan} and \emph{Prediction}). 
The model achieves qualitatively good prediction performance for individual subjects.
In addition, the remaining vertical panel of Fig.~\ref{fig:clinical_visualization_pred_dkt} shows the estimated abnormal \Ab trajectory \( \mathbf{b}_a(t) \) and the selected latent state trajectory \( \sum_k \mathbf{w}_{h}[k]\mathbf{h}_k(t) \) at five time points of disease-age \( t = 0, 0.25, 0.5, 0.75, 1 \). 
These results demonstrate that the model captures both subject-specific progression trajectories and common patterns at the cohort-level.
We perform the same experiments using the MUSE atlas, with results shown in Supplementary Fig.~\ref{fig:clinical_visualization_fitting_recon_muse_cn_mci_ad}, and the model performance is strong.

\begin{figure}[!htbp]
\centering
\includegraphics[width=1\textwidth]{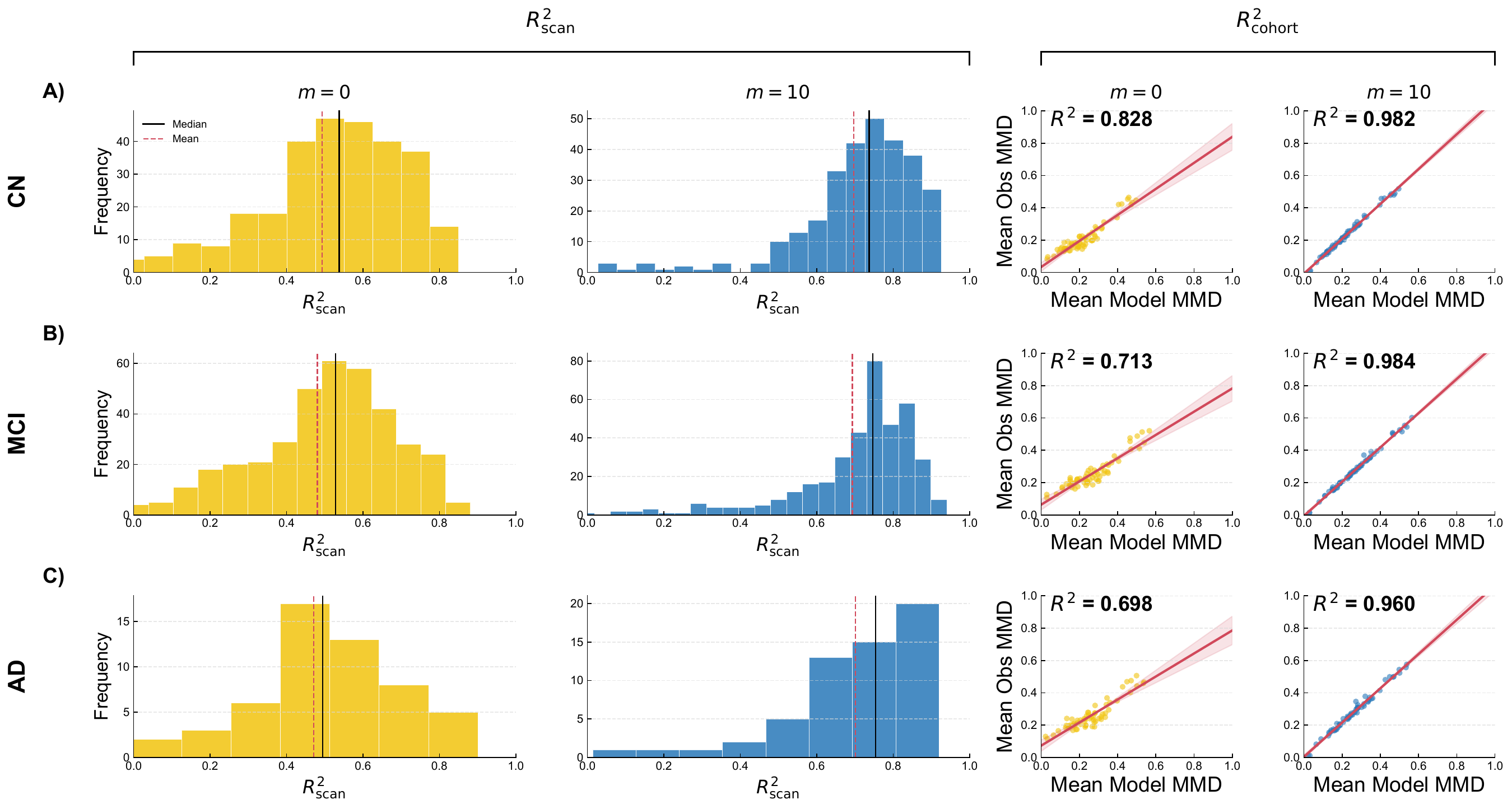}
\caption{
  \textbf{Prediction results comparison for ADNI dataset when $m=0$ and $m=10$ for CN, MCI and AD groups.}
  We perform the prediction experiment to predict the next scan for subjects with longitudinal data in ADNI dataset ({\bf `Pred1'} task).
  We compare the prediction performance when $m=0$ and $m=10$ for CN, MCI and AD groups in the metrics of $R_{\text{scan}}^2$ and $R_{\text{cohort}}^2$.
  From top to bottom, the three rows correspond to CN, MCI and AD groups cohort.
  From left to right, there are two panels showing the distribution of per-scan $R_{\text{scan}}^2$ scores and the cohort-level $R_{\text{cohort}}^2$ score.
  For each panel, we compare the prediction performance when $m=0$ and $m=10$.
}
\label{fig:prediction_performance_diff_groups_m0m10}
\end{figure}

\subsection{Cohort inversion maintains consistent performance across diagnosis groups.}
We further compare prediction performance (`Pred1' task) for the CN, MCI, and AD groups under the DKT atlas under two conditions: no latent states (\( m = 0 \)) and with ten latent states (\( m = 10 \)).
The results are shown in Fig.~\ref{fig:prediction_performance_diff_groups_m0m10}. 
Performance is evaluated using the per-scan \( R_{\text{scan}}^2 \) and the cohort-level \( R_{\text{cohort}}^2 \).
In the setting \( m = 0 \) (i.e., no latent states), the mean values \( R_{\text{scan}}^2 \) for the CN, MCI, and AD groups are 0.493, 0.481, and 0.471, respectively. 
In contrast, when \( m = 10 \), the mean values of \( R_{\text{scan}}^2 \) increase substantially to 0.697, 0.693, and 0.701 for the CN, MCI and AD groups, respectively. 
A similar trend is observed for the cohort-level metric. 
For \( m = 0 \), the mean values of \( R_{\text{cohort}}^2 \) are 0.828, 0.713, and 0.698 for the CN, MCI, and AD groups, respectively, while for \( m = 10 \), they improve to 0.982, 0.984, and 0.960.
These results demonstrate that the incorporation of latent states substantially improves prediction performance across diagnostic groups.

\begin{figure}[!htbp]
\centering
\includegraphics[width=1\textwidth]{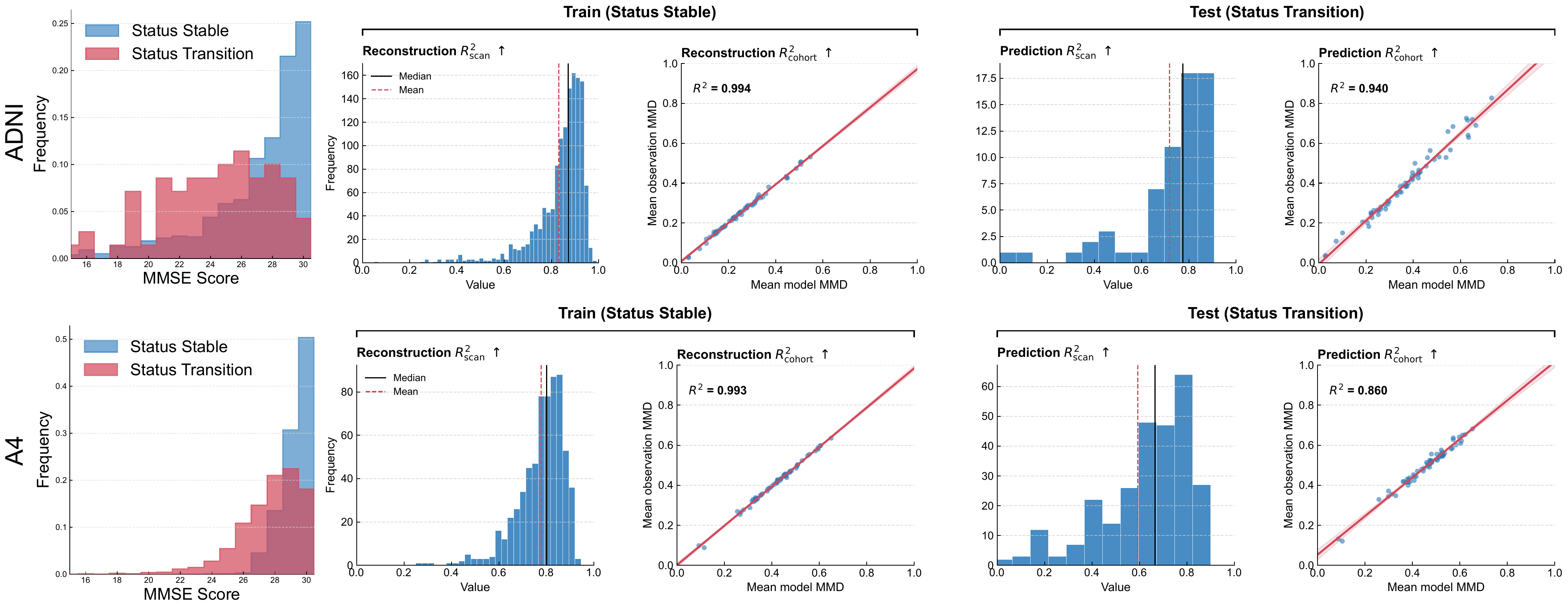}
\caption{
  \textbf{Prediction of unseen \Ab scans for ADNI and A4 subjects.}
  The trained simple binary classifier is applied to A4 Study subjects to identify individuals with potential diagnostic transitions from CN to AD. 
  For ADNI dataset, we use provided ground-truth longitudinal diagnosis labels to identify subjects with diagnostic transitions.
  For both datasets, we train the LNODE model using status-stable subjects and subsequently fix the cohort-shared parameters to predict \Ab test scans for subjects with potential status transitions ({\bf `Pred3'} task).
  The first row presents the prediction results for ADNI subjects, while the second row shows the corresponding results for A4 subjects. 
  From left to right, the panels display the histogram of MMSE scores for status-stable and status-transition subjects, the reconstruction accuracy during the training phase, and the prediction accuracy during the test phase, evaluated using the metrics \( R_{\text{scan}}^2 \) and \( R_{\text{cohort}}^2 \).
}
\label{fig:adni_a4_potential_transition_dkt}
\end{figure}

\begin{figure}[!htbp]
  \centering
  \includegraphics[width=1\textwidth]{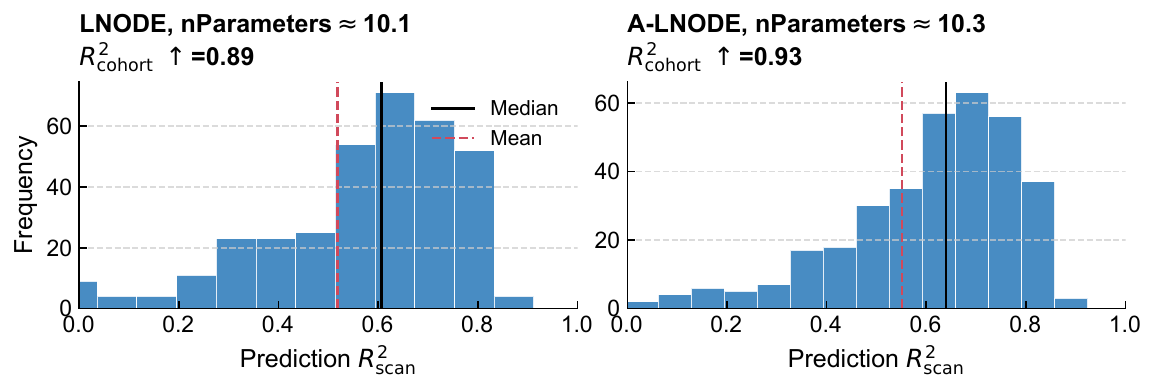}
  \caption{
    \textbf{Prediction of unseen \Ab scans for attention-based model.}
    We randomly shuffle the observation scans of the ADNI \Ab dataset and split the subjects into training and test sets, with 80\% of the subjects used for training and the remaining 20\% used for testing. 
    The results shown here correspond to the test set. 
    We compare the forecasting performance of the model with and without the learnable attention-based brain connectivity described in equation~\ref{eq:laplacian}. 
    During this experiment, we remove the bias term $\bm\phi_3$ and adopt the inversion scheme described in equation~\ref{e:inv2}.
    The results show that the LNODE model achieves lower cohort-level $R^2$ values (0.89 vs. 0.93) and lower scan-level $R^2$ values (0.51 vs. 0.55) compared with the attention-based model (A-LNODE). 
    These findings suggest that incorporating learnable brain connectivity improves the forecasting performance of the model.
    }
  \label{fig:alnode_experiment}
\end{figure}
 
\subsection{Accurate prediction of unseen scans of status transition.}
We further evaluate the generalization performance of the proposed LNODE model on unseen data from the ADNI and A4 cohorts. 
Both the ADNI and A4 datasets provide Mini-Mental State Examination (MMSE) scores, and ADNI additionally provides clinical diagnosis labels.
We divided subjects into two groups according to disease progression: 
(1) status-stable subjects whose clinical diagnosis remains unchanged on all available scans; 
(2) status-transition subjects whose clinical diagnosis changes during the observation period.
The subjects in the A4 Study are all cognitively normal (CN) at enrollment and do not have longitudinal diagnosis labels. 
To identify subjects with potential diagnostic transitions, we train a simple binary classifier using data from the ADNI cohort. 
Specifically, we collect MMSE scores from 1263 ADNI subjects and treat each MMSE score as an independent data point. 
A random forest classifier is trained using MMSE as the sole feature and achieves an accuracy of 98\% to distinguish CN from AD subjects. 
Hyperparameters of the classifier are selected via five-fold cross-validation. 
We then apply this classifier to the A4 cohort and identify 361 subjects with potential status transitions. 
The mean MMSE scores for A4 subjects with stable status and transition status are 29.23 and 27.56, respectively.
Then, subjects with stable diagnosis labels are used for training to estimate both subject-specific parameters and cohort-shared parameters. 
The cohort-shared parameters are then fixed, and only the subject-specific parameters are inverted using the first scan of subjects with potential status transitions.
\par
Figure~\ref{fig:adni_a4_potential_transition_dkt} (bottom row) presents the results for A4 subjects using the DKT atlas. 
For scan reconstruction, the per-scan \( R_{\text{scan}}^2 \) has a mean of 0.778, a median of 0.800, and a standard deviation of 0.098, while the cohort-level score \( R_{\text{cohort}}^2 \) is 0.993. 
For the prediction of unseen scans in subjects in transition status (`Pred3' task), \( R_{\text{scan}}^2 \) achieves a mean of 0.592, a median of 0.666 and a standard deviation of 0.282, with a cohort-level \( R_{\text{cohort}}^2 \) of 0.860. 
These results indicate a strong generalization of the parameters shared by the cohort to unseen subjects and datasets.
\par
The ADNI dataset provides longitudinal diagnosis information, from which 89 subjects are identified as exhibiting diagnostic transitions. 
The mean MMSE scores for status-stable and status-transition subjects are 26.50 and 22.36, respectively.
Figure~\ref{fig:adni_a4_potential_transition_dkt} (top row) presents the experimental results for ADNI subjects using the DKT atlas. 
In the reconstruction task, the per-scan \( R_{\text{scan}}^2 \) has a mean of 0.831, a median of 0.871, and a standard deviation of 0.147, while the cohort-level score \( R_{\text{cohort}}^2 \) is 0.994. 
For the prediction of unseen scans in status-transition subjects (`Pred3' task), \( R_{\text{scan}}^2 \) achieves a mean of 0.718, a median of 0.775 and a standard deviation of 0.205, with a cohort-level \( R_{\text{cohort}}^2 \) of 0.940.
Supplementary Fig.~\ref{fig:adni_a4_potential_transition_muse} presents the results of this experiment using the MUSE atlas. 
In general, these results demonstrate that the proposed model achieves an accurate prediction of unseen scans for subjects with potential diagnostic transitions across different datasets and atlases.
\par
We further investigate whether the attention-based learnable brain connectivity defined in equation~\ref{eq:laplacian} can improve the predictive performance of the model. 
To this end, we conduct the following experiment: the ADNI subjects are randomly divided into training (80\%) and test (20\%) sets. 
We train the model with and without equation~\ref{eq:laplacian} under the inversion scheme described in Equation~\ref{e:inv2}. 
We omit the term $\bm\phi_3$ of the model in this experiment.
The forecasting performance on the test set is presented in Fig.~\ref{fig:alnode_experiment}.
On the left, we show the performance of the LNODE model, which achieves a cohort-level $R^2_{\text{cohort}}=0.89$ and an average scan-level $R^2_{\text{scan}}=0.51$. 
On the right, we show the performance of the attention-based model (A-LNODE), which achieves $R^2_{\text{cohort}}=0.93$ and an average $R^2_{\text{scan}}=0.55$. 
These results suggest that the learnable brain connectivity improves the forecasting capability of the model on the ADNI dataset.
More important, A-LNODE only increase the model complexity from 10.1 parameters per subject to 10.3 parameters per subject.

\begin{figure}[!htbp]
\centering
\includegraphics[width=1\textwidth]{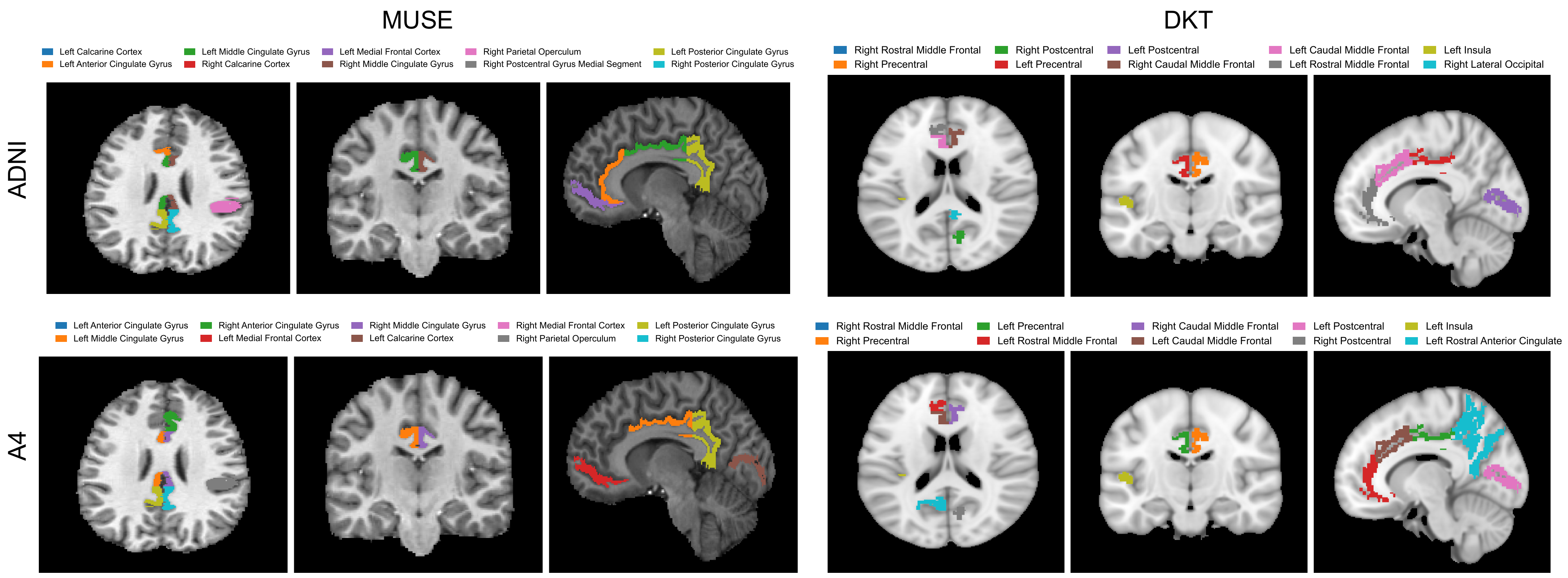}
\caption{
  \textbf{The most frequent \Ab initial condition (IC) locations identified by the algorithm for the MUSE and DKT atlases in the ADNI and A4 cohorts.}
  We perform model inversion to reconstruct \Ab scans for all subjects in the ADNI and A4 datasets using either the MUSE or DKT atlas. 
  From the inversion results, we extract the estimated initial condition (IC) locations and compute the occurrence frequency of each ROI across all subjects. 
 Here we show the ten most frequently identified IC locations.
  The two columns correspond to cases using the MUSE or DKT atlases, respectively, while the two rows correspond to cases using the ADNI or A4 datasets.
  Different atlases and datasets yield similar IC location patterns.
}
\label{fig:IC_loc_view_abeta}
\end{figure}

\subsection{Consistent \Ab initial seeding locations.}
In the LNODE model, the initial condition \( \mathbf{p}_i \) represents the initial seeding locations of abnormal \Ab at initial disease age. 
This quantity is an invertible parameter in the model inversion and is subject to a sparsity constraint to encourage only a small number of ROIs to act as initial seeds.
We analyze the common initial seed ROIs across subjects in both the ADNI and A4 datasets, using both the MUSE and DKT atlases. 
After model inversion, for each subject \( i \), we identify the nonzero elements of \( \mathbf{p}_i \) as the initial seed ROIs. 
We then compute the occurrence frequency of each ROI across all subjects. 
Fig.~\ref{fig:IC_loc_view_abeta} summarizes the ten most frequently selected ROIs for the ADNI and A4 datasets under both the MUSE and DKT atlases.
For the ADNI cohort using the DKT atlas, the ROI most frequently selected is the Right Rostral Middle Frontal region, with a selection probability of 0.639. 
Similarly, for the A4 cohort under the DKT atlas, the Right Rostral Middle Frontal region is the most frequently selected initial seed, with a selection probability of 0.760. 
Under the MUSE atlas, the most frequently selected ROI for the ADNI cohort is the Left Calcarine Cortex, with a selection probability of 0.687, whereas for the A4 cohort, the Left Anterior Cingulate Gyrus is most frequently selected, with a selection probability of 0.359.
Despite differences in atlas definitions and naming conventions, the selected initial seed regions exhibit substantial overlap across datasets and atlases. 
This consistency suggests that the initial seed locations inferred are robust and capture the common underlying anatomical patterns of the initiation of \Ab.

\begin{figure}[!htbp]
\centering
\includegraphics[width=1\textwidth]{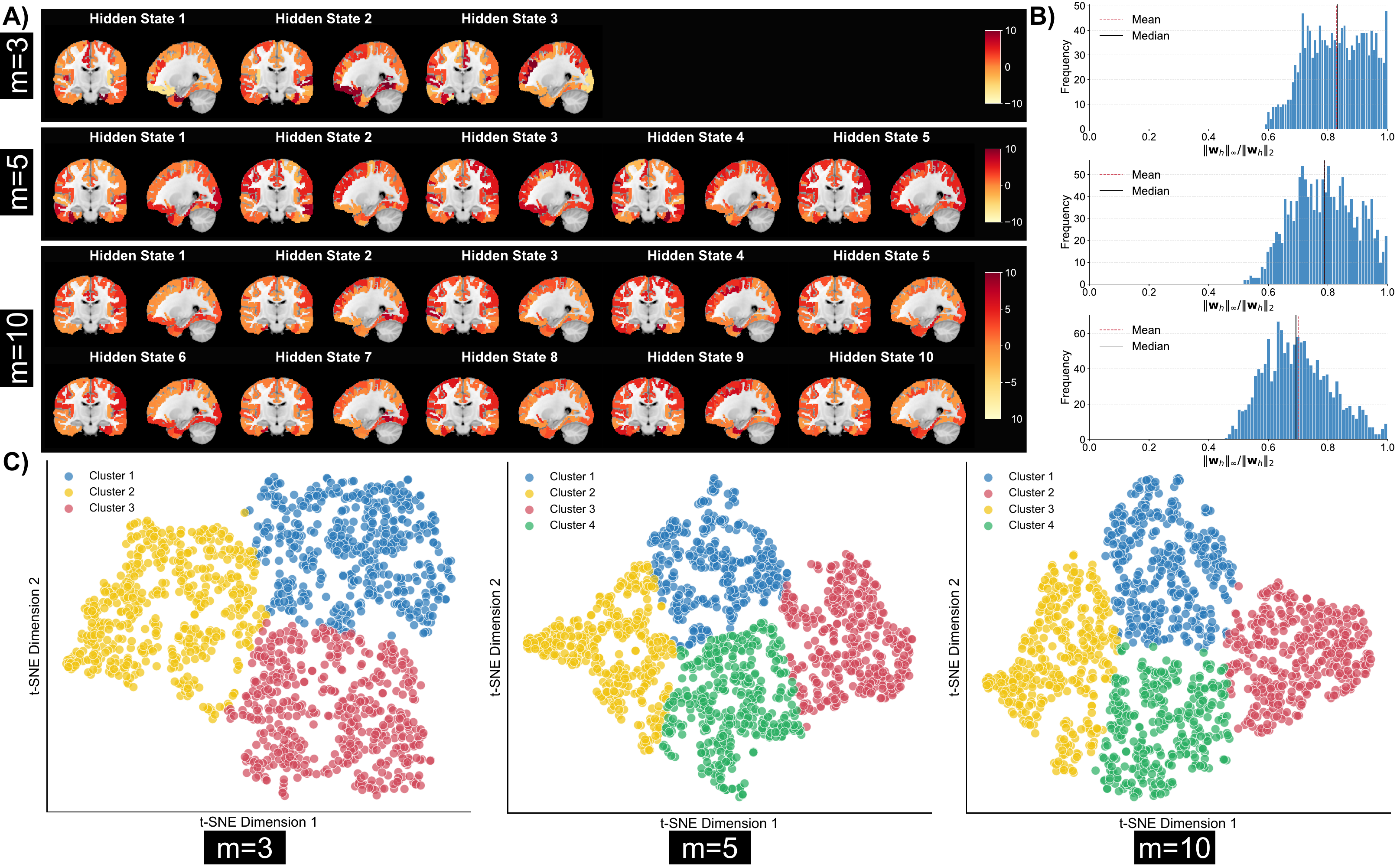}
\caption{
  \textbf{Latent states parameters analysis when $m=3, 5, 10$.}
  \textbf{A)} We show the cohort-shared latent state parameters \( \boldsymbol{\phi}_{3,k} \) for different numbers of latent states, \( m = 3, 5, 10 \). 
  \textbf{B)} For each latent state \( k \), we quantify the sparsity of the subject-specific weights \( \mathbf{w}_{h,i}[k] \) using the ratio \( \|\mathbf{w}_h\|_{\infty} / \|\mathbf{w}_h\|_{2} \). 
  Larger values of this ratio indicate greater sparsity. 
  This analysis is performed for the cases \( m = 3, 5, 10 \).
  \textbf{(C)} For a given choice of \( m \), for each subject \( i \), we compute the quantity \( \sum_{k} \mathbf{w}_h[k]\mathbf{h}_{i,k}(t = 0.5) \). 
  This \( N \)-dimensional representation is then embedded into a two-dimensional space using t-SNE. 
  We subsequently apply K-means clustering to the embedded representations and show the clustering results for the cases \( m = 3, 5, 10 \).
}
\label{fig:hidden_states_phi3}
\end{figure}

\begin{figure}[!htbp]
\centering
\includegraphics[width=1\textwidth]{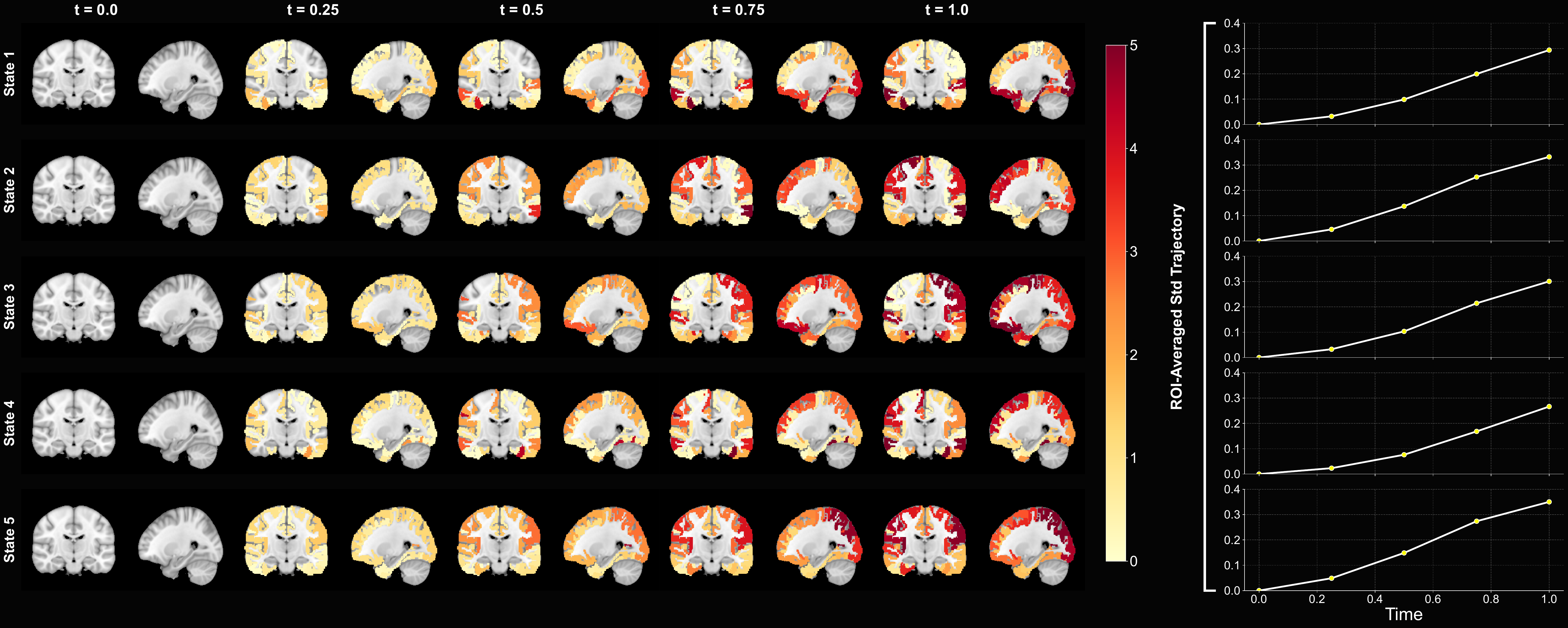}
\caption{
  \textbf{Latent state trajectory as a function of disease age.}
  We compute the mean and standard deviation of the latent state trajectories \( \mathbf{h}_{i,k}(t) \) across subjects \( i \) for different latent states. 
  From top to bottom, the five rows correspond to the five latent states under \( m = 5 \). 
  The first ten columns display the mean latent state trajectory at disease ages \( t = 0, 0.25, 0.5, 0.75, 1 \), with two brain views shown for each time point. 
  The final column presents the temporal evolution of the standard deviation of the latent state trajectories across subjects.
  At each disease age \( t \), the standard deviation is computed across subjects for each ROI and then averaged across all ROIs to yield a single standard deviation value. 
  Latent states are stable across subjects over disease age, with standard deviations remaining below 0.4 while the magnitude of the latent state lies in the range \([0, 5]\).
}
\label{fig:hidden_states_trajectories}
\end{figure}

\subsection{Latent states imply \Ab dynamic subtypes.}
We analyze the estimated latent states \( \mathbf{h}_{k,i}(t) \) of the LNODE model from two complementary perspectives: the cohort-shared bias term \( \boldsymbol{\phi}_{3,k} \) and the subject-specific latent state trajectories \( \mathbf{h}_{k,i}(t) \). 
First, we examine the latent state dynamics. 
Each latent state \( \mathbf{h}_{k, i}(t) \) is governed by three variables: \( \phi_{1,k} \), \( \phi_{2,k} \), and \( \boldsymbol{\phi}_{3,k} \), where \( \boldsymbol{\phi}_{3,k} \in \mathbb{R}^N \) is a constant source term. 
We test \( \boldsymbol{\phi}_{3,k} \) for different numbers of latent states \( m = 3, 5, 10 \) to investigate the spatial distribution of the latent states across ROIs. 
Figure~\ref{fig:hidden_states_phi3}(A) shows the estimated cohort-shared parameters \( \{\boldsymbol{\phi}_{3,k}\}_{k=1}^m \). 
The results show distinctive spatial patterns for each hidden state, which indicates different common trajectories of the disease spreading pathology.
\par
Second, we examine the sparsity level of the latent state selection weights \( \mathbf{w}_{h,i} \) across subjects by metric \(\|\mathbf{w}_{h,i}\|_\infty / \|\mathbf{w}_{h,i}\|_2\).
A value close to $1$ indicates that a single latent state dominates for subject $i$, whereas a value near $1/\sqrt{m}$ implies that multiple latent states contribute equally.
Larger values of this ratio indicate greater sparsity. 
As shown in Figure~\ref{fig:hidden_states_phi3}(B), the values of the ratio are significantly higher than the lower bound $1/\sqrt(m)$ in all latent states and the choices of \( m \), indicating that each subject predominantly selects only a few latent states.
\par
We map the quantity \( \sum_{k} \mathbf{w}_h[k]\mathbf{h}_{i,k}(t = 0.5) \) into a two-dimensional space using t-SNE~\citep{van2008visualizing}. 
K-means clustering is then applied to the resulting low-dimensional representations. 
The optimal numbers of clusters (3, 4, and 4) are determined using the Silhouette score for the cases \( m = 3, 5, 10 \), respectively. 
We examine the results at different disease age \(t=0.25, 0.5, 0.75, 1\).
Figure~\ref{fig:hidden_states_phi3}(C) shows the K-means clustering results.
These clustering results suggest the presence of potential subtypes of progression of \Ab. 
Notably, the patterns observed at \( t = 0.5 \) persist throughout the disease-age range \( t = 0.25, 0.5, 0.75, 1 \).
\par
In addition, for each subject \( i \), we compute the mean and standard deviation of the latent state trajectory \( \mathbf{h}_{i,k}(t) \) across disease age \( t \) for each latent state \( k \). 
The mean of the latent state dynamics characterizes common patterns of \Ab progression, while the standard deviation reflects the temporal stability of the latent state across disease age.
The estimated latent states exhibit strong stability across subjects. 
Figure~\ref{fig:hidden_states_trajectories} presents the mean trajectories of \( \mathbf{h}_{i,k}(t) \) across subjects \( i \) for the case \( m = 5 \). 
We further report the corresponding standard deviation of the trajectories to quantify inter-subject variability.
Across all latent states, the mean standard deviation averaged over ROIs increases gradually with disease age but remains below 0.4 throughout. 
Compared to the magnitude of approximately 5 observed in mean latent trajectories \( \mathbf{h}_{i,k}(t) \), this level of variability is relatively small. 
These results indicate that the bias term $\boldsymbol{\phi}_{3,k}$ dominates the latent state dynamics.

\section{Discussion}\label{s:discussion}
Spatiotemporal modeling of tau and amyloid-beta (\Ab) positron emission tomography (PET) imaging has been extensively explored in the literature~\citep{hong2022image,schafer2021bayesian,schafer2020network,schafer2021predicting,raj2025understanding,fan2024amyloidpetnet,jasodanand2025ai,vogel2021four}. Existing approaches span conventional statistical frameworks, deep learning architectures, and mechanism-based (e.g., biophysically informed) models.  
Classical statistical models are generally robust and interpretable; however, they are often restricted to characterizing global \Ab dynamics rather than capturing fine-grained regional or network-level processes. A representative example is the linear mixed-effects model used to characterize the longitudinal evolution of Alzheimer's disease biomarkers~\citep{gordon2018spatial}.  
Deep learning–based approaches, while powerful and flexible in modeling complex, high-dimensional imaging data, typically suffer from reduced interpretability and limited mechanistic transparency~\citep{jasodanand2025ai}.

Mechanism-based diffusion–reaction frameworks and their variants have been extensively investigated as mechanistic descriptions of the spatiotemporal propagation of pathological proteins in neurodegenerative disorders~\citep{iturria2014epidemic,vogel2020spread,wen2025single}, while more advanced formulations have been introduced to improve concordance with empirical observations~\citep{xu2025multiscale,thompson2020protein,pal2022nonlocal}. 
Nevertheless, the resulting parameterizations often remain predominantly subject-specific, which limits their capacity to represent population-level structure~\citep{vogel2021four,raj2025understanding}. 
Despite the intrinsic heterogeneity of Alzheimer's disease, converging evidence indicates the presence of reproducible, biologically meaningful, and shared
trajectories of disease progression across individuals~\citep{sintini2020longitudinal,lam2013clinical,whittington2018spatiotemporal,dong2017heterogeneity}. 
These observations motivate the explicit integration of cohort-level structure into models of disease progression.

Our proposed framework aligns closely with several converging directions in the existing literature. Cohort-level analyses have been increasingly recognized as a critical component in modeling the progression of neurodegenerative diseases. For instance, the Subtype and Stage Inference (SuStaIn) model estimates disease subtypes from cross-sectional and longitudinal imaging data by assigning individual scans to discrete disease stages and subtypes~\citep{young2018uncovering}. 
Subsequent work by Young \emph{et al.} further underscored the importance of cohort-level modeling for accurately characterizing longitudinal PET dynamics~\citep{young2024data}. In parallel, Bayesian inversion techniques have been applied to infer cohort-level priors from PET imaging data~\citep{chaggar2025personalised}. Although that study was limited to relatively small subsets of ADNI data, it demonstrated encouraging reconstruction performance and highlighted the value of incorporating spatiotemporal information at the cohort-level. 
Collectively, these efforts illustrate that models explicitly capturing shared cohort dynamics are becoming central to advancing our understanding of \Ab and tau progression. Within this context, the adoption of cohort-shared parameters in the present work constitutes a promising strategy to address persistent challenges in modeling the dynamics of neurodegenerative diseases.

In this context, LNODE performs the estimation of a unified mechanism-based model from large-scale \Ab PET datasets acquired from both the ADNI and A4 cohorts. 
The inferred cohort-level shared parameters characterize common spatiotemporal patterns of \Ab propagation across individuals, whereas subject-specific parameters capture residual inter-individual variability. This joint modeling strategy yields substantial gains in both retrospective scan reconstruction accuracy and prospective prediction performance. In particular, we show that even when fixed pretrained cohort parameters are applied to new subjects, LNODE preserves high predictive accuracy on previously unseen PET scans, indicating that these shared parameters encode robust and transferable disease-relevant dynamics.

LNODE incorporates a sparsely selected set of latent states, which is critical to the overall performance of the model. 
Within this framework, cohort-shared parameters that govern the dynamics of the latent states play a central role in capturing common progression patterns that cannot be explained solely by classical diffusion and reaction processes. The latent states can be interpreted as additional source terms that represent shared but unobserved influences on \Ab{} dynamics. 
These latent contributions may reflect the aggregate effects of multiple biological and environmental determinants implicated in Alzheimer’s disease (AD) pathophysiology, including genetic susceptibility, vascular comorbidities, and metabolic dysregulation~\citep{gottesman2017association,morris2010apoe}. 
By embedding these effects within a structured dynamical system, the model provides an interpretable extension of traditional biophysical frameworks while preserving the capacity to capture cohort-level regularities in \Ab propagation.

An aspect that is sometimes overlooked is the numerical stability of an image analysis framework. In fact, the intrinsic inaccuracy and ill-posedness of the underlying inverse problem frequently constrain the reliability of the inferred parameters and dynamical trajectories, as discussed in previous studies ~\citep{liu2024parameter,browning2024structural}. To overcome this robustness limitation, we introduced multiple sparsity based regularization terms. We systematically assess the stability of the proposed inversion algorithm, including a series of synthetic experiments (Fig.~\ref{fig:synthetic_results}). In these evaluations, LNODE accurately reconstructs both the \Ab\ scans and the corresponding ground-truth parameters.

In the clinical datasets, the quality of both reconstruction and prediction, as well as inferred latent states, remains consistently robust between individuals, as shown in Fig.~\ref{fig:clinical_quantitative_inversion_results} and Fig.~\ref{fig:hidden_states_trajectories}. 
A more detailed examination of the latent states reveals reproducible dynamical patterns shared across subjects (Fig.~\ref{fig:hidden_states_phi3}). 
Together, these findings underscore that a stable and well-posed inversion framework is crucial for deriving reliable biophysical inferences from spatiotemporal PET imaging data.  As an additional demonstration, Fig.~\ref{fig:adni_a4_potential_transition_dkt} illustrates an experiment in which cohort-level parameters are estimated using subjects with stable diagnoses and then transferred to subjects undergoing diagnostic transitions. 
In this setting, the model accurately forecasts previously unseen \Ab PET scans. 
The mean interval between consecutive \Ab PET acquisitions is 2.34~(\(\pm\)~0.94) years in the ADNI cohort and 4.68~(\(\pm\)~0.93) years in the A4 Study.
Despite these relatively long inter-scan intervals, the proposed framework yields accurate predictions that are suitable for supporting downstream clinical decision-making. Beyond predictive performance, the model also offers mechanistic insight into disease evolution. 
As shown in Fig.~\ref{fig:IC_loc_view_abeta}, the inferred initial seeds of abnormal \Ab consistently concentrate within a small, recurrent subset of brain regions across datasets and parcellation schemes. 
The convergence toward coherent spatiotemporal propagation patterns, in conjunction with robustly identified initial seeding locations, provides a data-driven characterization of the mechanisms governing \Ab dissemination and may guide the design and spatial targeting of future therapeutic interventions.

LNODE exhibits several limitations.  
From a modeling standpoint, the latent states are permutation invariant: any reordering of the latent states, together with their associated sparse selection weights, yields solutions that are mathematically equivalent for the underlying system of ordinary differential equations.  
This identifiability issue arises from the absence of explicit regularization terms or ordering constraints imposed on the latent states.  
Incorporating additional regularization strategies or structural priors could promote more interpretable latent representations and facilitate a clearer separation of CN, MCI, and AD subjects into well-defined latent states.  
Addressing these modeling challenges constitutes an important direction for future research.

Furthermore, the present study focuses exclusively on \Ab-PET data.  
Extending the framework to jointly model \Ab, tau, and structural atrophy within a unified architecture is warranted.  
Such an extension would enable a more comprehensive characterization of AD progression and would support a more rigorous evaluation of the generalizability of the proposed approach across multiple imaging biomarkers.  

Finally, although the latent states are informative, they remain abstract constructs and may capture a complex mixture of confounding biological and environmental influences.  
A more precise biological interpretation of the cohort-shared parameters and the associated latent dynamics will be necessary to strengthen mechanistic insight and to support more reliable early diagnosis and therapeutic intervention strategies.

\section{Conclusion}
We have introduced a novel, biophysically grounded framework for modeling \Ab PET signal dynamics. 
The proposed model achieves high-fidelity reconstruction of observed scans and exhibits robust predictive performance on held-out datasets. 
Moreover, the favorable stability properties of the associated inversion solver furnish a principled and reliable foundation for parameter estimation in biophysical models of Alzheimer's disease-related biomarkers. 
The inferred common initial seeding patterns, in conjunction with the characterization of latent-state dynamics, indicate shared progression trajectories across individuals. 
Future work will focus on extending this framework to incorporate additional imaging markers, cerebrospinal fluid \Ab and tau measurements, as well as demographic and clinical variables, with the overarching aim of attaining a more comprehensive and mechanistically informed representation of Alzheimer's disease progression.

\section{Data availability}
Alzheimer's Disease Neuroimaging Initiative (ADNI) database can be accessed via \url{https://adni.loni.usc.edu/}.
Anti-Amyloid Treatment in Asymptomatic Alzheimer's Disease (A4) Study database can be accessed via \url{https://www.a4studydata.org}.

\section{Code availability}
Python implementation of the algorithm is available on the Github page: \url{https://github.com/Zheyu-Wen/LNODE_code}.

\section{Funding}
This material is based upon work supported by NSF award OAC 22042261; 
by the U.S. Department of Energy, Office of Science, Office of Advanced Scientific Computing Research, Applied Mathematics program, Mathematical Multifaceted Integrated Capability Centers (MMICCS) program, under award number DE-SC0023171; 
and by the U.S. National Institute on Aging under award number R21AG074276-01. 
Any opinions, findings, and conclusions or recommendations expressed herein are those of the authors and do not necessarily reflect the views of the DOE, NIH, and NSF. 
Computing time on the Texas Advanced Computing Centers Stampede system was provided by an allocation from TACC and the NSF. 

\section{Acknowledgements}\label{s:acknowledgements}
Data used in preparation of this article were obtained from the Alzheimer's Disease Neuroimaging Initiative (ADNI) database (adni.loni.usc.edu). 
As such, the investigators within the ADNI contributed to the design and implementation of ADNI and/or provided data but did not participate in analysis or writing of this report. 
A complete listing of ADNI investigators can be found at: \url{https://adni.loni.usc.edu/wp-content/uploads/how_to_apply/ADNI_Acknowledgement_List.pdf}.
The A4 and LEARN Study Leadership Teams provide the list below for standardized acknowledgement of the many individuals who have contributed to the A4 and LEARN Studies (current and former): 
\url{https://www.actcinfo.org/wp-content/uploads/2024/07/A4-LEARN-Study-Team-List-Longitudinal-as-of-May-2023-Journal-Version-1.pdf}

\section{Competing interests}\label{s:competing_interests}
The authors declare that they have no known competing financial interests or personal relationships that could have appeared to influence the work reported in this paper.


\bibliographystyle{plainnat}

\bibliography{cas-refs}

\appendix
\section{Appendix}
\renewcommand{\thefigure}{A\arabic{figure}}
\setcounter{figure}{0}

\begin{figure}[!htbp]
\centering
\includegraphics[width=1\textwidth]{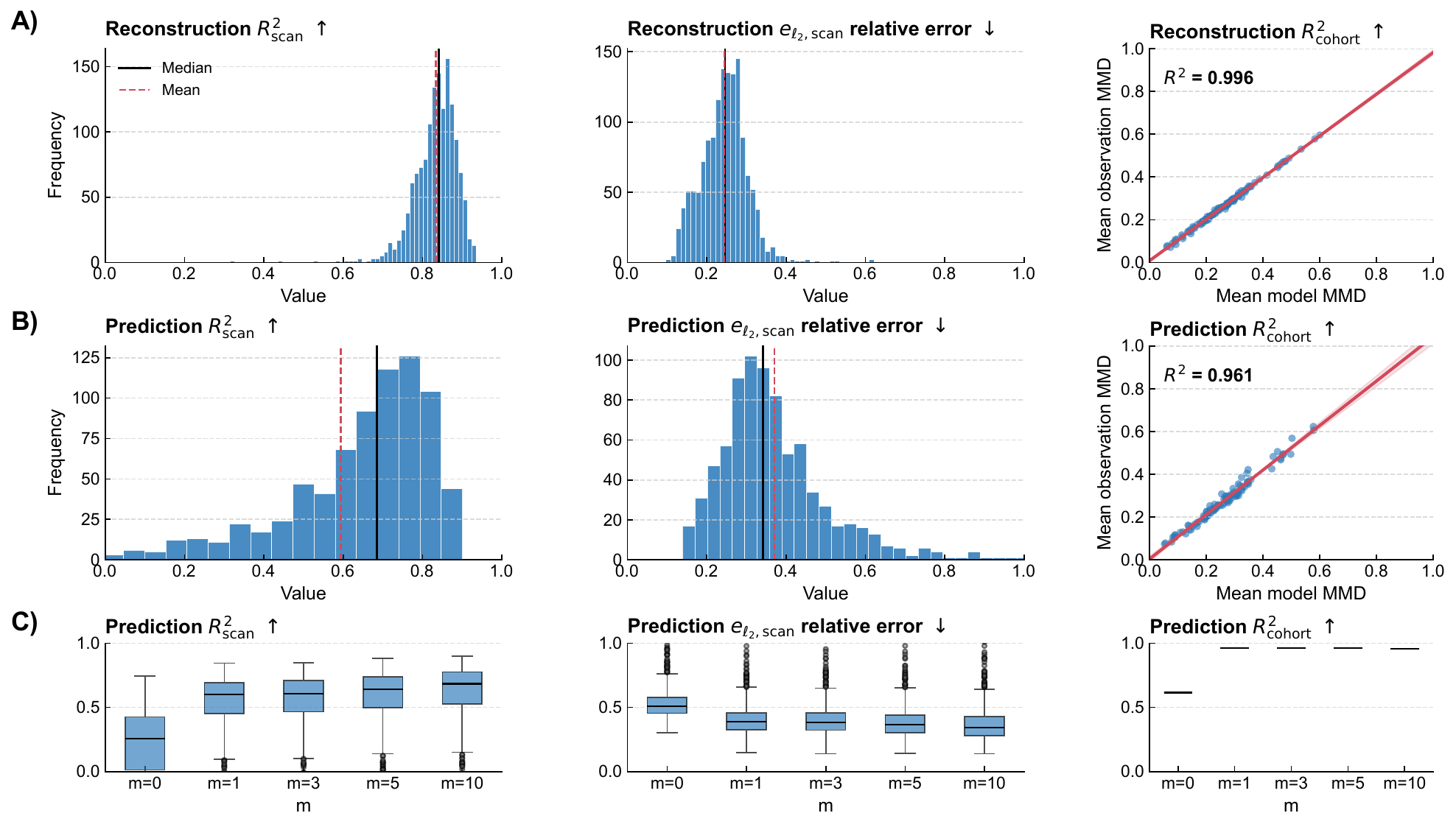}
\caption{
  \textbf{Quantitative evaluation of reconstruction and prediction using the MUSE atlas.}
  We assess model performance on the ADNI cohort through two tasks: reconstruction using all scans and prediction of the next scan using only the first scan for parameter estimation. 
  The top row shows reconstruction with $m=10$, the middle row shows next-scan prediction with $m=10$, and the bottom row compares prediction performance across $m=0,1,3,5,10$. 
  From left to right, the panels report the distributions of per-scan $R_{\text{scan}}^2$, per-scan relative error $e_{\text{scan}}$, and the cohort-level $R_{\text{cohort}}^2$.
  With $m=10$, reconstruction achieves a mean $R_{\text{scan}}^2$ of 0.834 (median 0.841, s.d.\ 0.053), while prediction attains a mean of 0.688 (median 0.740, s.d.\ 0.194). 
  Increasing the number of latent states consistently improves prediction accuracy. 
  Without latent states ($m=0$), the mean $R_{\text{scan}}^2$ is 0.276 (median 0.326, s.d.\ 0.273), which increases substantially to 0.605 (median 0.651, s.d.\ 0.195) for $m=1$. 
  At the cohort level, $R_{\text{cohort}}^2$ rises from 0.600 for $m=0$ to 0.966 for $m=10$.
  Across diagnosis groups, mean $R_{\text{scan}}^2$ values for CN, MCI, and AD improve from 0.334, 0.264, and 0.056 at $m=0$ to 0.699, 0.684, and 0.619 at $m=10$. 
  Similarly, $R_{\text{cohort}}^2$ increases from 0.494, 0.167, and $-0.614$ to 0.961, 0.958, and 0.899, respectively.
}
\label{fig:inversion_results_muse}
\end{figure}

\begin{figure}[!htbp]
\centering
\includegraphics[width=1\textwidth]{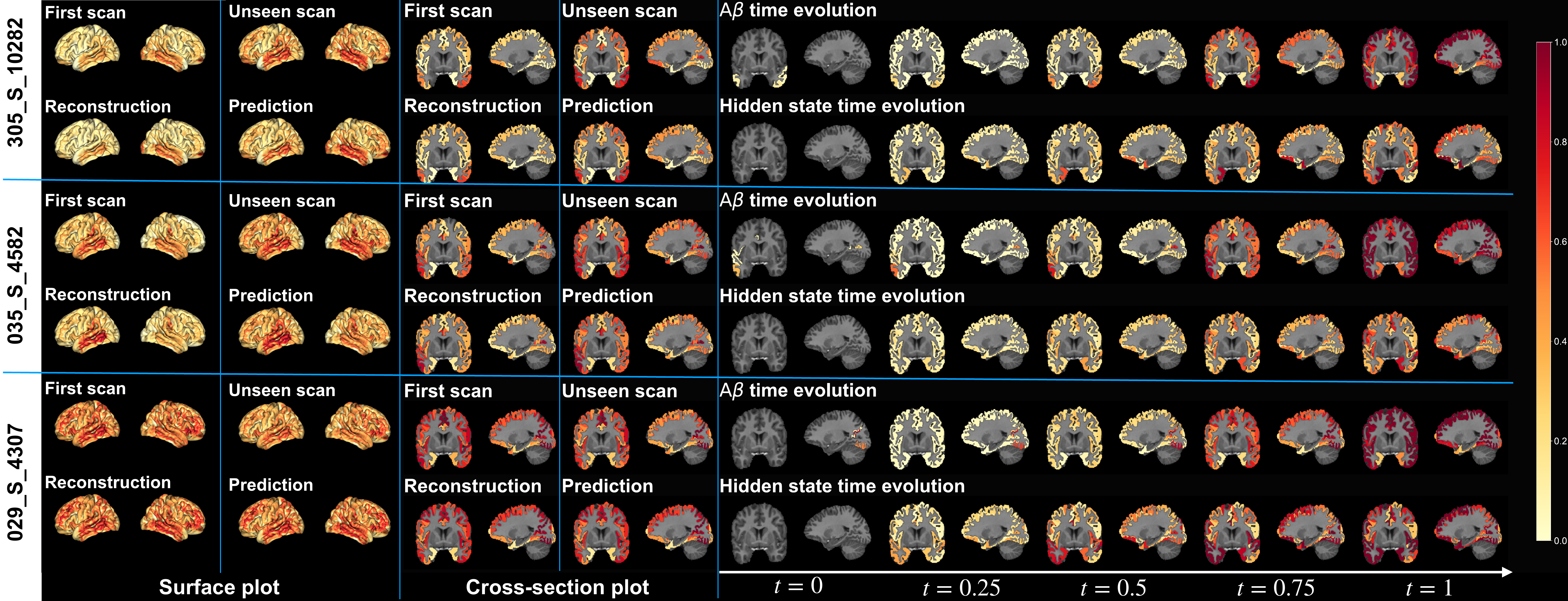}
\caption{
  \textbf{Qualitative evaluation of the model reconstruction and prediction performance using the MUSE atlas.} 
  We perform model inversion using the first scan of each subject as input and predict the next scan for subjects in the ADNI cohort and we use MUSE atlas here. 
  From top to bottom, the three rows correspond to three representative subjects, selected from the CN, MCI, and AD groups, respectively. 
  These subjects exhibit substantial changes in \Ab burden between the first and next scans.
  From left to right, the figure contains five panels: surface plots of reconstruction performance, surface plots of prediction performance, cross-section plots of reconstruction performance, cross-section plots of prediction performance, and the estimated trajectories of \Ab burden and latent states across disease age \( t \). 
  Each of the first four panels spans two columns to display two different brain views. 
  The final panel spans ten columns, illustrating the estimated abnormal \Ab trajectory \( \mathbf{b}_a(t) \) and the selected latent state trajectory \( \sum_k \mathbf{w}_{h}[k]\mathbf{h}_k(t) \) at five disease-age time points \( t = 0, 0.25, 0.5, 0.75, 1 \), with two brain views shown at each time point.
  Overall, the model successfully reconstructs and predicts qualitatively consistent \Ab burden patterns for individual subjects. 
  The estimated trajectory plots further provide information about the interaction between the latent states and subject-specific \Ab progression.
}
\label{fig:clinical_visualization_fitting_recon_muse_cn_mci_ad}
\end{figure}

\begin{figure}[!htbp]
\centering
\includegraphics[width=1\textwidth]{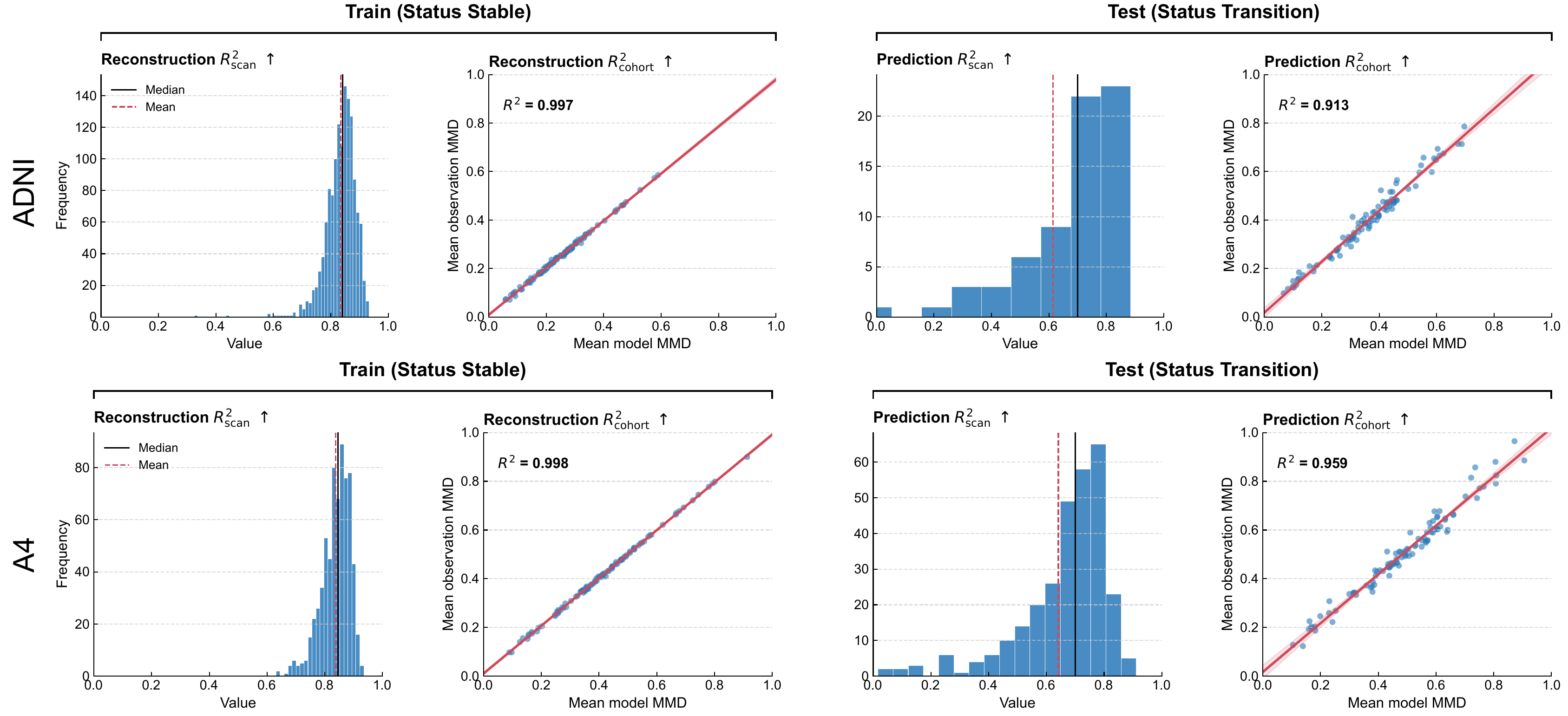}
\caption{
  \textbf{Prediction of unseen \Ab scans for ADNI and A4 subjects in MUSE atlas.}
  The trained binary classifier is applied to A4 Study subjects to identify individuals with potential diagnostic transitions from CN to AD. 
  The ADNI dataset contains ground-truth diagnosis transition information.
  For both datasets, we train the LNODE model using diagnosis-stable subjects and subsequently fix the cohort-shared parameters to predict \Ab test scans for subjects with potential diagnosis transitions.
  The first row presents the prediction results for ADNI subjects, while the second row shows the corresponding results for A4 subjects. 
  From left to right, the panels display the reconstruction accuracy during the training phase, and the prediction accuracy during the test phase, evaluated using the metrics \( R_{\text{scan}}^2 \) and \( R_{\text{cohort}}^2 \).
  For A4 subjects, reconstruction achieves a mean $R_{\text{scan}}^2$ of 0.838 (median 0.845, s.d.\ 0.049) and $R_{\text{cohort}}^2 = 0.998$, while prediction on diagnosis-transition subjects yields a mean $R_{\text{scan}}^2$ of 0.641 (median 0.700, s.d.\ 0.216) and $R_{\text{cohort}}^2 = 0.959$.  
  For ADNI, reconstruction attains a mean $R_{\text{scan}}^2$ of 0.835 (median 0.842, s.d.\ 0.052) with $R_{\text{cohort}}^2 = 0.997$, and prediction on diagnosis-transition subjects reaches a mean $R_{\text{scan}}^2$ of 0.614 (median 0.700, s.d.\ 0.321) with $R_{\text{cohort}}^2 = 0.913$.
}
\label{fig:adni_a4_potential_transition_muse}
\end{figure}

\end{document}